%% file: ms.tex
\documentclass[11pt]{article}
\usepackage[T1]{fontenc}
\usepackage[utf8]{inputenc}
\usepackage[a4paper,margin=2.5cm]{geometry}
\usepackage{lmodern}
\usepackage{microtype}
\usepackage{setspace}
\usepackage{amsmath,amssymb}
\usepackage{graphicx}
\usepackage{booktabs}
\usepackage{array}
\usepackage{longtable}
\usepackage{textcomp}
\usepackage[font=small,labelfont=bf]{caption}
\usepackage[natbibapa]{apacite}
\usepackage{xcolor}
\usepackage[hidelinks]{hyperref}
\graphicspath{{figures/}}

\newcommand{\fun}[1]{\texttt{#1}}
\begin{document}
\onehalfspacing

\begin{center}
{\Large\bfseries Fitting Bayesian stable isotope mixing models in under a second with prior-data fitted networks}\par\vspace{1em}
{\large Andrew~C.~Parnell$^{1}$, Richard~Inger$^{2}$, Brice~X.~Semmens$^{3}$, Eric~J.~Ward$^{4}$, Andrew~L.~Jackson$^{5}$ and Brian~C.~Stock$^{6}$}\par\vspace{0.8em}
\end{center}
{\small
$^{1}$School of Mathematics and Statistics and UCD Earth Institute, University College Dublin, Dublin, Ireland\\
$^{2}$Centre for Ecology and Conservation, University of Exeter, Penryn, Cornwall, UK\\
$^{3}$Scripps Institution of Oceanography, University of California San Diego, La Jolla, California, USA\\
$^{4}$Northwest Fisheries Science Center, NOAA Fisheries, Seattle, Washington, USA\\
$^{5}$Zoology, School of Natural Sciences, Trinity College Dublin, Dublin, Ireland\\
$^{6}$Institute of Marine Research, Norway\par\vspace{0.5em}
\textbf{Correspondence:} Andrew C. Parnell, \href{mailto:andrew.parnell1@ucd.ie}{andrew.parnell1@ucd.ie}\par}

\section*{Abstract}

\begin{enumerate}
\item Bayesian mixing models, most widely through the MixSIAR software, are the standard way to estimate diet and source proportions from stable isotopes and other biotracers. MixSIAR fits each model by Markov chain Monte Carlo (MCMC), which takes minutes to hours and often fails to converge at standard run lengths. This limits how many models, datasets and sensitivity analyses a study can explore.
\item We present MixSIARPFN, which replaces MixSIAR's MCMC with prior-data fitted networks: transformer networks trained once on millions of datasets simulated from MixSIAR's own model and priors. For a new dataset, they return the posterior distribution of every diet proportion and model parameter in a single pass. The networks cover MixSIAR's model class, and an R package keeps MixSIAR's input files and function calls.
\item On 400 simulated datasets with known diets, held out from all development, MixSIARPFN was 3\% to 4\% less accurate than MCMC on average for MixSIAR's usual error structures, and its 95\% intervals contained the true diet as often (96\% to 97\% of the time). Each fit took under a second on one processor core, over 3,000 times faster than MixSIAR. When simulated data broke MixSIAR's assumptions, MixSIARPFN lost accuracy no faster than MCMC, except with heavy-tailed or skewed errors in one design (up to 13\% less accurate). It matched MixSIAR closely on five of MixSIAR's nine real examples. It differed more on the geese and wolves examples, but MixSIAR's own model fits these data poorly, and changing MixSIAR's error structure moved its answer as much as switching to MixSIARPFN. A sensitivity analysis of 3,600 fits took an hour instead of about half a year.
\item MixSIARPFN makes Bayesian mixing models practical where MCMC is not, so that users can fit every candidate model, analyse many datasets and test sensitivity to discrimination factors and priors. We recommend confirming the final model with MCMC where the data fit the model poorly or conclusions rest on small differences.
\end{enumerate}

\textbf{Data and code:} the R package (\url{https://github.com/andrewcparnell/MixSIARPFN}), the trained networks (\url{https://doi.org/10.5281/zenodo.24002247}) and the code and results for this paper are openly available (Data availability statement). 

\textbf{Key-words:} amortised inference, Bayesian mixing models, diet estimation, MixSIAR, neural networks, prior-data fitted networks, simulation-based inference, stable isotopes 

\section{Introduction}

Biotracer mixing models estimate the proportions of sources in a mixture, most often the contributions of prey to a consumer's diet, from the tracer values of the consumers and the sources \citep{phillips2003isosource, parnell2010siar, phillips2014best}. Bayesian versions propagate the uncertainty in source values and trophic discrimination, and let the diet vary with consumer groups and covariates \citep{moore2008mixsir, semmens2009hierarchical, ward2010source, parnell2013bayesian}. MixSIAR \citep{stock2018mixsiar} brought these developments together and added a choice of error structures \citep{stock2016error}. It is now among the most used tools in isotope ecology, and is applied well beyond diet studies, for example to apportion sediment and pollution sources \citep{blake2018dmixsiar, torresmartinez2021nitrate, wang2024phosphorus}.

MixSIAR fits each model by Markov chain Monte Carlo (MCMC) in JAGS \citep{plummer2003jags}. A single fit takes from minutes to hours, and the chains must then be checked for convergence \citep{gelman1992inference, vehtari2021rank}. With nested random effects or weakly identified sources, MixSIAR's standard run lengths often fail and must be rerun longer. The cost multiplies whenever a study fits many models: comparing covariates or error structures, analysing many populations, testing tracer combinations \citep[see][for examples]{chalauxclergue2024tracer, liang2026artificial}, or checking sensitivity to discrimination factors and priors \citep{bond2011sensitive, swan2019evaluating}. Faster alternatives exist, but each still fits every dataset from scratch. The simmr and cosimmr packages use variational approximations \citep{govan2023simmr, govan2024cosimmr}, and MixSIAPy has recently reimplemented MixSIAR's model with the no-U-turn sampler \citep{yang2026mixsiapy}.

Amortised inference removes the per-dataset computation altogether \citep{cranmer2020frontier, zammitmangion2025neural}. A neural network is trained once, on datasets simulated from the model with parameters drawn from the prior, to return the posterior given the data, so a new dataset needs only a single pass through the network. This neural posterior estimation is now well established \citep{papamakarios2016fast, lueckmann2017flexible, greenberg2019automatic, lueckmann2021benchmarking, radev2022bayesflow}. Prior-data fitted networks \citep[PFNs;][]{muller2022transformers} are transformers trained to minimise a proper scoring rule, so that the best network outputs the posterior itself. Their tabular form, TabPFN \citep{hollmann2023tabpfn, hollmann2025accurate}, reads a whole dataset as a table, and transformers of this kind have recently been shown to perform posterior inference in context \citep{mittal2025incontext, reuter2025can, vetter2025effortless}.

Networks trained on simulations already infer population-genetic and phylodynamic parameters \citep{sheehan2016deep, flagel2019unreasonable, voznica2022deep} and past biodiversity \citep{cooper2025deepdiver}, and reviews have argued for their wider use in ecology \citep{borowiec2022deep, pichler2023machine}. Their accuracy must be checked carefully, however, because a network's posterior can be overconfident in ways that are hard to see without a reference \citep{hermans2022trust}. Amortised inference has recently been extended to hierarchical models \citep{habermann2025amortized}, including a single network that serves a whole family of generalised linear mixed models \citep{kipnis2026metabeta}, and applied to one environmental source-apportionment problem \citep{altin2026heavymetal}. To our knowledge it has not been applied to biotracer mixing models, whose simplex-valued proportions, source and discrimination uncertainty and error structures make them a distinctive target.

\begin{figure}[t]
\centering 
\includegraphics[width=\textwidth]{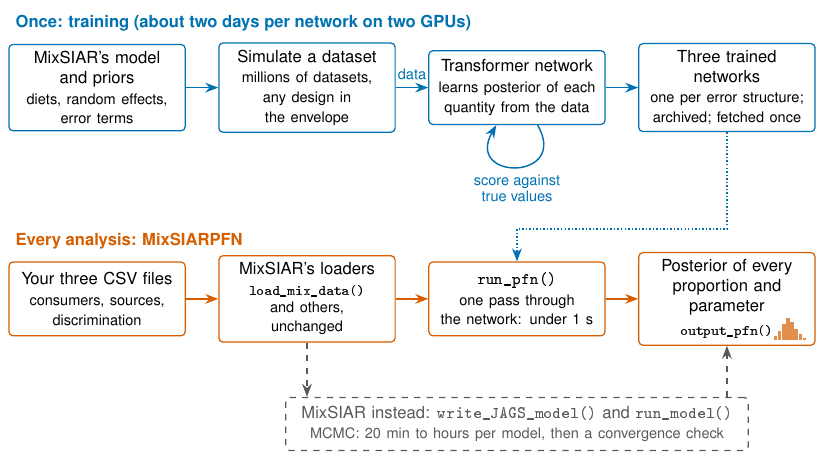}
\caption{How MixSIARPFN works. For each of MixSIAR's three error structures, a transformer network is trained once on millions of datasets simulated from MixSIAR's model and priors, across every design the package supports, and learns to return the posterior of each diet proportion and parameter from the data. Every analysis then reads the user's files with MixSIAR's own loaders, and a single pass through the network (under a second) gives the posteriors that MixSIAR would obtain by Markov chain Monte Carlo (MCMC) sampling over minutes to hours.}
\label{fig:schematic}
\end{figure}

Here we present MixSIARPFN (Fig.~\ref{fig:schematic}): three prior-data fitted networks, one per MixSIAR error structure, that together cover the full MixSIAR model class, and an R package that reads MixSIAR's input files with MixSIAR's own function calls. It replaces MixSIAR's inference without changing its inputs. We ask whether the networks are close enough to MCMC in accuracy and calibration to be used in its place, at a small fraction of the cost. We test them on held-out data simulated from MixSIAR's model, on simulated data that break the model and on MixSIAR's nine real examples, one of which we use for a sensitivity analysis that would be impractical with MCMC.

\section{Materials and methods}

\subsection{MixSIAR's model}

We first write out the simplest MixSIAR model, with one diet for all consumers, and then describe how it is extended \citep[see][for the full model]{stock2018mixsiar}. Let $\mathbf{y}_i = (y_{i1},\dots,y_{iJ})$ be the values of $J$ tracers (for example $\delta^{13}$C and $\delta^{15}$N) for consumer $i = 1,\dots,N$, and let $\mathbf{p} = (p_1,\dots,p_K)$ be the proportions of $K$ sources in the diet, with $p_k \geq 0$ and $\sum_k p_k = 1$. For each source $k$ and tracer $j$, the data give the sample mean $m_{kj}$, standard deviation $s_{kj}$ and size $n_k$ of the source values, the mean $\lambda_{kj}$ and standard deviation $\tau_{kj}$ of the trophic discrimination factor (the shift in tracer value between diet and consumer) and, optionally, the source's elemental concentration $c_{kj}$. The source means $\mu_{kj}$ and variances $\omega^2_{kj}$ are estimated with the diet:
\[
\mu_{kj} \sim \mathrm{N}\!\left(m_{kj},\, s^2_{kj}/n_k\right), \qquad \frac{(n_k - 1)\, s^2_{kj}}{\omega^2_{kj}} \sim \chi^2_{n_k}, \qquad \mathbf{p} \sim \mathrm{Dirichlet}(\alpha_1,\dots,\alpha_K).
\]
A consumer's expected tracer value is the mixture of the discrimination-corrected source means, 
\[
\eta_j = \frac{\sum_k p_k\, c_{kj}\, (\mu_{kj} + \lambda_{kj})}{\sum_k p_k\, c_{kj}},
\]
with all $c_{kj} = 1$ unless the model has concentration dependence \citep{phillips2002concentration}. The likelihood is normal, $y_{ij} \sim \mathrm{N}(\eta_j, \sigma^2_j)$, independently across consumers, and MixSIAR's three error structures \citep{stock2016error} differ only in the variance $\sigma^2_j$. Under \emph{Process only} error, the variance comes entirely from the variability of the sources and of discrimination, propagated through the diet:
\[
\sigma^2_j = v_j(\mathbf{p}) = \textstyle\sum_k p_k^2 \left(\omega^2_{kj} + \tau^2_{kj}\right).
\]
This structure has no free variance parameter and suits a consumer that samples its prey at random. Under \emph{Residual only} error, $\sigma^2_j$ is instead a free parameter (a covariance matrix across tracers, with a Wishart prior) that absorbs all variation among consumers, whatever its cause. Under \emph{Residual~$\times$~Process} error, MixSIAR's default, $\sigma^2_j = \xi_j\, v_j(\mathbf{p})$ with $\xi_j \sim \mathrm{U}(0, 20)$, so the process variance is scaled up or down by a term per tracer. The choice matters for inference: Process only error ties the spread of the consumers to the diet itself, so posteriors narrow quickly with more consumers, whereas the residual terms let the data set the spread. We therefore trained one network per error structure.

MixSIAR extends this model in three ways, all of which MixSIARPFN supports. First, the diet can vary among consumers: the isometric log-ratio transform \citep{egozcue2003ilr} of consumer $i$'s diet is the global diet's plus the effects of the consumer's levels of up to two factors (fixed or random, crossed or nested) and a slope times a continuous covariate, with $\mathrm{U}(0, 20)$ priors on random-effect standard deviations. Second, source values may differ between the levels of one factor, for example by region. Third, source data can be supplied as raw samples, which MixSIAR models jointly with the diet. Appendix S1 gives these extensions in full.

\subsection{Amortised inference with prior-data fitted networks}

A prior-data fitted network $q_\phi(\theta \mid \mathcal{D})$ is trained on pairs $(\theta, \mathcal{D})$, where $\theta$ is drawn from the prior and the dataset $\mathcal{D}$ is simulated from the model given $\theta$, to minimise the expected negative log score $-\log q_\phi(\theta \mid \mathcal{D})$. Because the log score is a strictly proper scoring rule \citep{gneiting2007strictly}, the minimiser is the posterior $p(\theta \mid \mathcal{D})$ for every dataset the prior can produce \citep{muller2022transformers, nagler2023statistical}. We ask each network for marginal posteriors, with one query per diet proportion (global, per level, along the covariate, or for any combination of levels and covariate values) and one per parameter. Each answer is a bar distribution, a histogram over 1,000 bins: 500 evenly spaced and 500 even on the logit scale, which resolves posteriors close to 0 or 1 \citep{muller2022transformers}. Such histograms can take any shape, including the skewed and multimodal marginals that mixing models often produce.

\subsection{The networks and their training}

Each network reads a dataset as a table, following TabPFN \citep{hollmann2025accurate}. The columns are the tracers (and the covariate, if there is one), and the rows are of three kinds. Consumer rows hold the consumers' tracer values. Source rows, one per source and source-factor level, hold seven numbers per tracer: the discrimination-corrected mean, standard deviation, standard error, discrimination standard deviation, sample size, concentration and Dirichlet prior weight. Query rows are copies of their source's row, tagged with what is asked.

Factor levels enter as embeddings drawn at random from a pool of 20, so the network treats levels as exchangeable, as MixSIAR's model does. At use, answers are averaged over eight random draws. Eight transformer layers \citep{vaswani2017attention} alternate attention across columns within each row and across rows within each column, with queries attending to the data but not to each other \citep{kossen2021self}. Rows have no positions, so the answer is exactly invariant to the order of consumers and of sources (Appendix S2). Each network has 17.4 million parameters.

Training datasets were simulated afresh at every step from MixSIAR's model and priors, over every design MixSIARPFN supports: 2 to 8 sources, 1 to 4 tracers, 1 to 1,000 consumers, up to two effects of 2 to 20 levels, an optional covariate, concentration dependence, sources by factor and Dirichlet priors from very concentrated to very diffuse (Appendix S1). Each network was trained for 450,000 steps of 131,072 table cells on two NVIDIA L40S GPUs (about two days), using schedule-free AdamW \citep{defazio2024road}. The released checkpoint of each network is the one with the lowest loss on 24,000 fixed validation datasets never used for training (Appendix S2).

The released Residual~$\times$~Process network was trained for a further 150,000 steps with an additional task: predicting each diet one source at a time, as a share of what the earlier sources leave (autoregressive stick-breaking). This made its ordinary marginal posteriors, which the package uses, slightly more accurate. Of the other changes we tried during development, only longer training clearly improved accuracy (Appendix S9). The networks are exported to ONNX \citep{onnxruntime} and archived with a DOI.

\subsection{The MixSIARPFN package}

MixSIARPFN reimplements MixSIAR's loaders (\fun{load\_mix\_data()}, \fun{load\_source\_data()} and \fun{load\_discr\_data()}) with the same arguments and checks, so existing MixSIAR scripts need only one change: \fun{run\_pfn()} in place of \fun{write\_JAGS\_model()} and \fun{run\_model()}. Its \fun{output\_pfn()} replaces \fun{output\_JAGS()}, producing MixSIAR's summary statistics and posterior plots, and \fun{compare\_mixsiar()} compares a fit with MixSIAR's own. Raw source samples are summarised to means, standard deviations and sample sizes. This differs from MixSIAR's raw-data model, which estimates each source's distribution, including the correlation between tracers, jointly with the diets (Appendix S5).

Inference runs in Python through reticulate \citep{kalinowski2026reticulate}, but users need no Python: the first fit installs a small private environment (about 180~MB) and downloads the network it needs (about 70~MB), after which a fit takes about a second. Designs outside the networks' training envelope (Appendix S1) are rejected with a message. The package returns marginal posteriors only, so it has no counterpart to MixSIAR's \fun{combine\_sources()}, contrasts between groups, model comparison by leave-one-out cross-validation (\fun{compare\_models()}) or posterior predictive checks, all of which need the joint posterior (Table~\ref{tab:features}).

\begin{table}[t]
\caption{What MixSIARPFN provides compared with MixSIAR. MixSIARPFN returns the marginal posterior of every quantity MixSIAR reports, so the summaries and plots of a MixSIAR analysis are available, but analyses that need the joint posterior are not yet.}
\label{tab:features}
\centering 
\footnotesize 
\begin{tabular}{@{}p{5.2cm}p{9.6cm}@{}}
\toprule
MixSIAR analysis & MixSIARPFN \\
\midrule
Posterior of each diet proportion, standard deviation, residual term and slope; summaries and plots & Yes: marginal posteriors, through \fun{output\_pfn()} \\
Sums and differences of proportions: \fun{combine\_sources()}, contrasts between groups & No: needs joint draws (joint-diet training, Appendix S6, is a first step) \\
Model comparison by leave-one-out cross-validation (\fun{compare\_models()}) & No: needs pointwise likelihoods under the joint posterior \\
Posterior predictive checks of the fitted model & No: needs joint draws; run MixSIAR on the final model \\
Raw source data modelled jointly with the diets & Approximated: raw samples are summarised to means, standard deviations and sample sizes \\
Convergence diagnostics as a per-fit warning & None needed, but also no per-fit check of the approximation; \fun{compare\_mixsiar()} compares with a MixSIAR fit \\
\bottomrule
\end{tabular}
\end{table}

\subsection{Evaluation}

We evaluated MixSIARPFN in four steps, each answering one question (Table~\ref{tab:evaluations}). Where the data were simulated, the true diet was known and we scored each method against it. On MixSIAR's real examples, where it is not, we measured how far MixSIARPFN's answer was from MixSIAR's.

The first step asks how accurate and well calibrated MixSIARPFN is when MixSIAR's model is true. We copied the designs of two of MixSIAR's examples exactly: geese \citep[251 consumers in eight groups as a fixed effect, four sources, concentration dependence;][]{inger2006geese, stock2016mixsiarpkg} and wolves (66 consumers in eight packs nested in three regions as random effects, three sources given by region). For each design, we simulated 200,000 datasets from MixSIAR's prior and kept the 100 whose standardised group means and standard deviations were nearest the real data, so that the test datasets resemble real data.

Three test sets used MixSIAR's usual error structures: geese-like with Residual only error, as in MixSIAR's example; wolves-like with Residual~$\times$~Process error and random-effect standard deviations from MixSIAR's prior, which pushes many pack diets close to a single source; and wolves-like with realistic standard deviations (uniform on 0.2 to 2). A fourth set, geese-like with Process only error, tests the third error structure. Because earlier batches of these designs guided development, the test sets were simulated with new seeds only after the networks were fixed, under a protocol committed beforehand, and scored once (Appendix S3). Beyond these two designs, we applied simulation-based calibration \citep{talts2018validating, modrak2025simulation} to 2,000 datasets per network from the training distribution (Appendix S2), and compared MixSIARPFN with converged MixSIAR on 40 further designs (Appendix S4).

We fitted each dataset with MixSIARPFN and with MixSIAR 3.1.12 at its ``normal'' and ``long'' run lengths (100,000 and 300,000 iterations, three chains). We also fitted MixSIAPy 0.1.0 \citep[a preprint;][]{yang2026mixsiapy}, which samples MixSIAR's model with the no-U-turn sampler, and, on the geese-like sets, cosimmr 1.0.12 \citep{govan2024cosimmr}, which fits a simpler model by variational Bayes \citep{tran2021tutorial}. Our main reference is MixSIAPy, the best available approximation to the exact posterior on every dataset. It passed the current convergence rule \citep[rank-normalised $\hat{R}$ below 1.01 and effective sample sizes above 400;][]{vehtari2021rank} on 89\% to 100\% of the datasets of the three usual designs, whereas MixSIAR's long runs almost never did. We also report the pre-registered comparison, with MixSIAR's long runs where their largest Gelman--Rubin diagnostic was at most 1.1 \citep{gelman1992inference}.

The second step asks how MixSIARPFN behaves when the data break MixSIAR's model, as real data do. Under a second protocol, committed before its data were generated, we simulated geese-like and realistic wolves-like datasets as before and perturbed them in one of five ways, 20 datasets each. The perturbations were heavy-tailed errors (Student-$t$ with 3 degrees of freedom, rescaled to the model's variance), skewed errors (skew-normal, shape 5), outlying consumers (5\% moved by 4 standard deviations in one tracer), biased discrimination factors (data generated with means one standard deviation, and at least 0.5\textperthousand, above the values given to the methods) and overstated source precision (sample sizes reported five times too large). We fitted these with MixSIARPFN, MixSIAPy and MixSIAR (Appendix S8). We also fitted all nine of MixSIAR's isotope examples with each example's settings (16 models, including the eight alligator models of MixSIAR's example script) and compared them with converged MixSIAR runs, with posterior predictive checks of MixSIAR's own fits for three of them (Appendix S5).

The third step asks how fast MixSIARPFN is: each method fitted five datasets per test design on an otherwise idle machine, with MixSIARPFN running through its R package on one processor core. The fourth asks what the speed makes possible. Conclusions from a mixing model depend on choices the data cannot settle, above all the discrimination factors and the model structure, and checking them means refitting the model many times \citep{bond2011sensitive, phillips2014best}. We therefore refitted MixSIAR's alligator example \citep{stock2018mixsiar} with MixSIARPFN under every combination of its eight model structures, the three error structures, discrimination factors shifted by $-1$ to $+1$\textperthousand\ for each tracer in steps of 0.5, discrimination standard deviations halved, unchanged or doubled, and an uninformative or an illustrative informative prior (mean 75\% freshwater prey, worth four observations). This gave 3,600 fits, and we recorded which of the original study's conclusions held across them.

We used three measures. Accuracy, where the truth is known, is the continuous ranked probability score \citep[CRPS;][]{matheson1976scoring, gneiting2007strictly} of the posterior distribution $F$ of a diet proportion against its true value $p$, 
\[
\mathrm{CRPS}(F, p) = \int_0^1 \big(F(x) - \mathrm{H}(x - p)\big)^2\, dx = \mathrm{E}|X - p| - \tfrac{1}{2}\,\mathrm{E}|X - X'|,
\]
where $\mathrm{H}$ is the Heaviside step function ($\mathrm{H}(z) = 1$ for $z \geq 0$ and 0 otherwise) and $X$ and $X'$ are independent draws from $F$. The CRPS penalises posteriors that are off-centre and those that are needlessly wide, and is zero only for a point mass at the truth. We computed it from posterior draws.

To compare two methods, we averaged each method's CRPS over the diet proportions of a dataset that both report, took the ratio of the two methods' means over datasets, and report it as a percentage difference. Its uncertainty comes from the Bayesian bootstrap \citep{rubin1981bayesian}: each of 4,000 draws gives the datasets random $\mathrm{Dirichlet}(1,\dots,1)$ weights and recomputes the ratio, and the 2.5\% and 97.5\% quantiles of the draws give a 95\% credible interval. Both methods' scores on a dataset share its weight, so the comparison is paired.

Calibration is the coverage of 95\% intervals, that is, the share that contain the true proportion, which should be 95\%. True values within $4.6\times10^{-5}$ of 0 or 1, beyond the networks' finest histogram bin, were counted as 0 or 1. Where the truth is unknown, agreement with MixSIAR is the absolute difference between the two methods' posterior means, in percentage points of diet.

\begin{table}[t]
\caption{The four steps of the evaluation: for each, the data, whether the true diet was known, the methods MixSIARPFN was compared with, the measure, and where the results are. The held-out and misspecified datasets were simulated only after the networks were fixed, under protocols committed in advance.}
\label{tab:evaluations}
\centering 
\footnotesize 
\begin{tabular}{@{}p{3.3cm}p{4.3cm}p{1.1cm}p{2.6cm}p{2.0cm}p{1.3cm}@{}}
\toprule
Question & Data & Truth known & Compared with & Measure & Results \\
\midrule
Accurate when MixSIAR's model is true? & 400 held-out simulated datasets (geese-like and wolves-like designs); simulation-based calibration across the supported model space & Yes & MixSIAPy, MixSIAR, cosimmr & CRPS, coverage & Section~3.1, Table~\ref{tab:heldout}, Fig.~\ref{fig:accuracy} \\
Robust when the data break the model? & 200 held-out simulated datasets with perturbed data; MixSIAR's nine real examples & Simulated: yes; real: no & MixSIAPy, MixSIAR & CRPS, coverage; difference in posterior means & Section~3.2, Fig.~\ref{fig:misspec}, Appendix S5 \\
How fast? & Five held-out datasets per test design & n/a & All & Time per fit & Section~3.3, Fig.~\ref{fig:speed} \\
What does speed make possible? & MixSIAR's alligator example, 3,600 fits & No & None & Conclusions that hold & Section~3.4, Fig.~\ref{fig:sweep} \\
\bottomrule
\end{tabular}
\end{table}

\section{Results}

\subsection{Accuracy and calibration when MixSIAR's model is true}

\begin{figure}[t]
\centering 
\includegraphics[width=\textwidth]{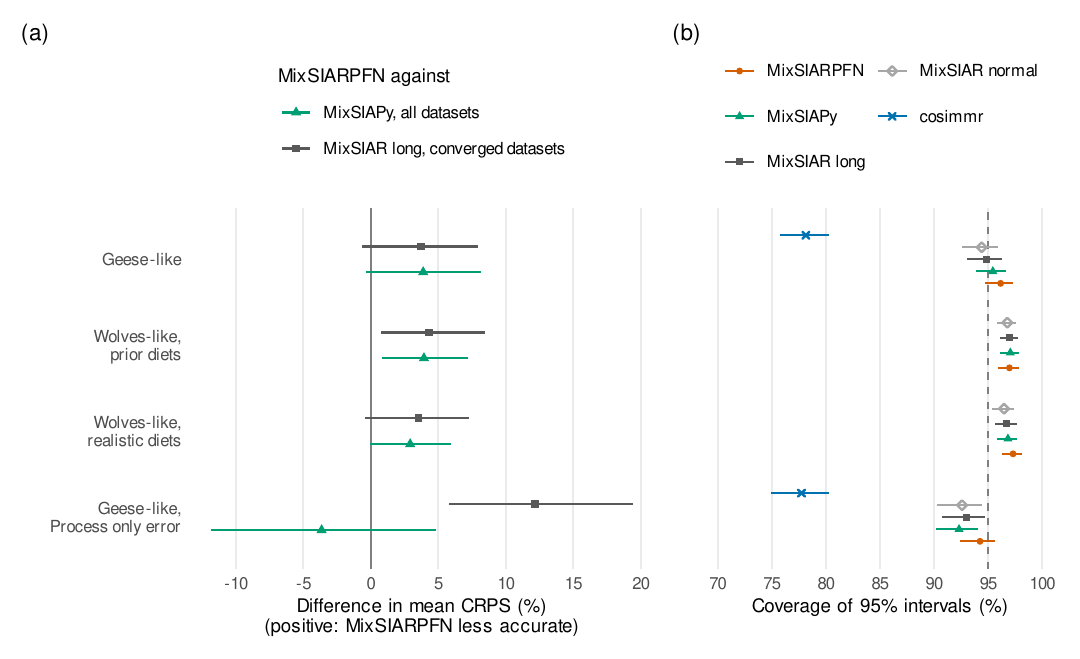}
\caption{Accuracy and calibration on the held-out test designs (100 simulated datasets each). (a) Difference in mean continuous ranked probability score (CRPS) between MixSIARPFN and Markov chain Monte Carlo (MCMC), with paired Bayesian bootstrap 95\% credible intervals over datasets; MCMC is MixSIAPy on all datasets, and MixSIAR's long runs on the datasets where they converged (the pre-registered comparison), and positive values mean MixSIARPFN is less accurate. (b) Share of 95\% intervals containing the true diet proportion, with Bayesian bootstrap 95\% credible intervals over datasets; the dashed line marks 95\%. With MixSIAR's usual error structures, MixSIARPFN was 3\% to 4\% less accurate than MCMC and as well calibrated, whereas cosimmr, which fits a simpler model, under-covered.}
\label{fig:accuracy}
\end{figure}

On the three test designs with MixSIAR's usual error structures, MixSIARPFN was slightly less accurate than MCMC (Table~\ref{tab:heldout}, Fig.~\ref{fig:accuracy}a). Its mean CRPS was 2.9\% to 3.9\% higher than MixSIAPy's over all 100 datasets per design, and 3.5\% to 4.3\% higher than that of converged MixSIAR long runs in the pre-registered comparison, with 95\% intervals reaching from about 0\% to 8\%. Even so, MixSIARPFN was the more accurate method on 32\% to 38\% of individual datasets. In diet terms, its posterior means differed from MixSIAPy's by a median of 0.9 to 1.4 percentage points, and by more than 10 points for 2\% to 3\% of proportions.

MixSIARPFN was as well calibrated as MCMC: its 95\% intervals contained the true proportion 96\% to 97\% of the time, against 95\% to 97\% for MixSIAPy and MixSIAR (Fig.~\ref{fig:accuracy}b). Its intervals were 4\% to 9\% wider than MixSIAPy's, so the networks err on the side of caution, and part of their small loss of accuracy is this extra width. MixSIAPy and converged MixSIAR agreed to within 0.7\%. cosimmr, which fits a simpler model than the one that generated the data, was 77\% to 220\% less accurate and covered the truth only 78\% to 79\% of the time.

With Process only error, MixSIARPFN was as accurate as MCMC over all 100 datasets (CRPS 3.6\% lower than MixSIAPy's, 95\% credible interval $-11.9\%$ to $+4.8\%$) and better calibrated (94\% coverage, against 92\% to 93\%). This average, however, combines two groups. On the datasets where MCMC converged, MixSIARPFN was 8\% to 12\% less accurate than MCMC, but on the 37 where MixSIAPy did not, it was 15\% more accurate than both samplers, which struggled there. The Process only network is thus the least accurate of the three where MCMC can be trusted.

Across the whole supported model space, the networks were close to calibrated. In simulation-based calibration of 2,000 datasets per network, the coverage of 50\%, 90\% and 95\% intervals was 49\% to 51\%, 90\% to 91\% and 94\% to 96\% for every kind of quantity (Appendix S2). On 40 further designs spanning the supported envelope, MixSIARPFN's posterior means differed from converged MixSIAR's by a median of 1.0 percentage point, and both methods' 95\% intervals covered 97\% of the true values (Appendix S4).

\begin{table}[t]
\caption{Accuracy and calibration of MixSIARPFN on the four held-out test designs (100 simulated datasets each, with known true diets). CRPS: percentage difference in mean continuous ranked probability score from the reference method, with a paired Bayesian bootstrap 95\% credible interval over datasets; positive means MixSIARPFN is less accurate. Better: share of datasets on which MixSIARPFN was the more accurate. The pre-registered comparison with converged MixSIAR uses only the datasets whose long run converged (number in brackets). Coverage: share of 95\% intervals containing the true proportion, all datasets (MixSIAPy's in brackets). Width: difference in mean 95\% interval width from MixSIAPy's. With MixSIAR's usual error structures, MixSIARPFN was 3\% to 4\% less accurate than MCMC, with well-calibrated, slightly wider intervals.}
\label{tab:heldout}
\centering 
\footnotesize 
\resizebox{\textwidth}{!}{\input{tables/table1_heldout.tex}}
\end{table}

\subsection{When the data break MixSIAR's model}

\begin{figure}[t]
\centering 
\includegraphics[width=\textwidth]{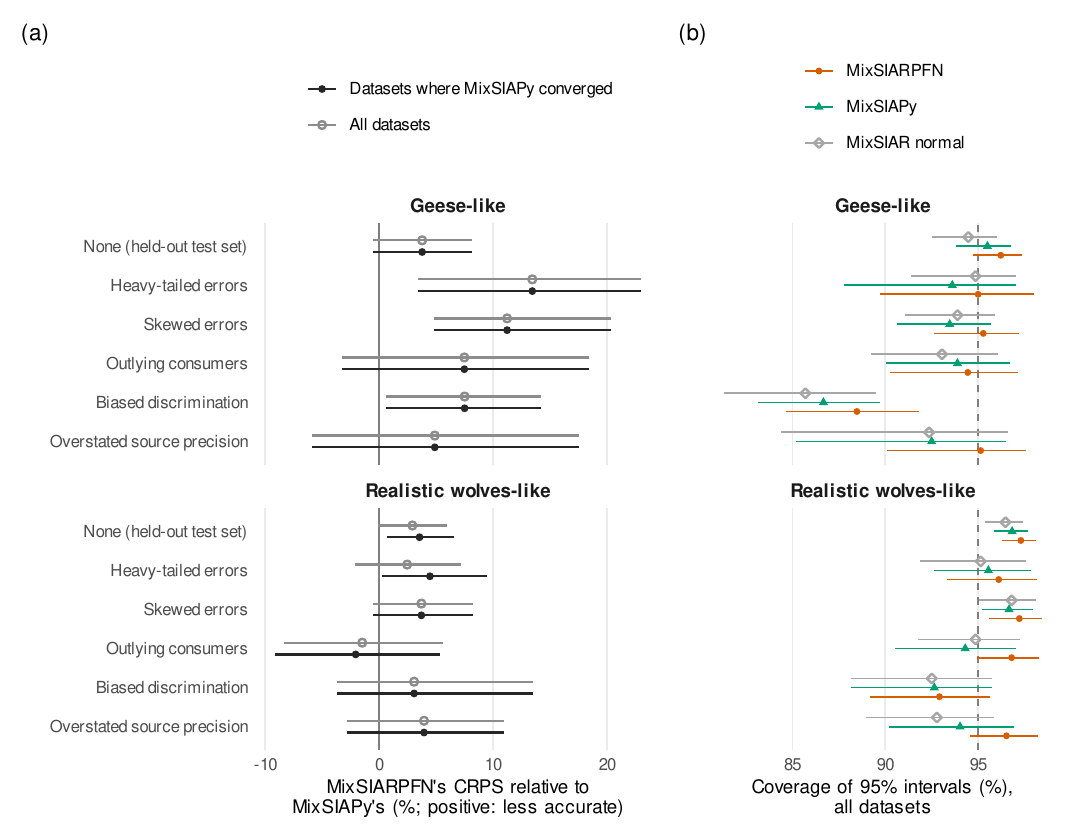}
\caption{MixSIARPFN on misspecified data: datasets of the geese-like and realistic wolves-like test designs whose consumers or inputs break MixSIAR's assumptions (20 per perturbation), compared with the clean held-out test sets (100 each). (a) Difference in mean continuous ranked probability score (CRPS) between MixSIARPFN and MixSIAPy, a Markov chain Monte Carlo (MCMC) sampler of MixSIAR's model, with paired Bayesian bootstrap 95\% credible intervals over datasets; positive values mean MixSIARPFN is less accurate. (b) Coverage of 95\% intervals for each method, with Bayesian bootstrap 95\% credible intervals; the dashed line marks 95\%. MixSIARPFN degraded about as much as MCMC, except under heavy-tailed and skewed errors on geese-like data, and stayed as well calibrated. Biased discrimination factors reduced every method's coverage.}
\label{fig:misspec}
\end{figure}

When simulated data broke MixSIAR's assumptions, MixSIARPFN's accuracy relative to MixSIAPy fell on average by 2.1 percentage points (95\% credible interval $-1.5$ to $+5.6$; Fig.~\ref{fig:misspec}). On the wolves-like design it fell no more than MCMC's under any perturbation, staying between 2\% below and 4.5\% above MixSIAPy, against 3.5\% above on unperturbed data. On the geese-like design, with 251 consumers, heavy-tailed and skewed errors increased the gap from 4\% to 13\% and 11\%, and outlying consumers and biased discrimination factors increased it to 7.5\%. With only 20 datasets per perturbation, the intervals exclude zero only for heavy-tailed errors, skewed errors and biased discrimination factors. Calibration held up, with MixSIARPFN covering the truth as often as MCMC under every perturbation. Biased discrimination factors reduced every method's coverage alike (to 86\% to 89\% on geese-like data), which is a failure of the model rather than of the inference.

On MixSIAR's nine real examples the true diet is unknown, so we compared MixSIARPFN with MixSIAR. The two agreed closely on five: every diet proportion of the killer whale, mantis shrimp, snail and storm petrel examples, and 98\% of those of the eight alligator models, was within 5 percentage points of MixSIAR's posterior mean. On the other four, 58\% to 67\% of proportions were within 5 points. Two of these, lake and Palmyra, have raw source data, which MixSIARPFN summarises, so the two methods fit slightly different models (Appendix S5).

The geese and wolves examples need more care, because MixSIAR's own model fits these data poorly. Its posterior predictive checks put 11 of 40 group statistics for geese, and 5 of 40 for wolves, outside the range of data generated by the fitted model (Appendix S5). On these examples, single proportions differed from MixSIAR's by up to 27 and 22 points, in the split of the diet between sources that the data barely distinguish. The two methods estimate the sum of these sources to within 4 points.

With a model that does not describe the data, MixSIAR's answer is not a reliable reference. On geese, changing MixSIAR's error structure alone moves its answer as much as switching to MixSIARPFN does, and MixSIARPFN's answer is close to MixSIAR's under that other error structure (Appendix S5). Repeated MCMC runs agree closely with each other, so the differences come from the methods rather than from MCMC noise, and both datasets lie outside what the networks were trained on.

\subsection{Speed}

\begin{figure}[t]
\centering 
\includegraphics[width=0.92\textwidth]{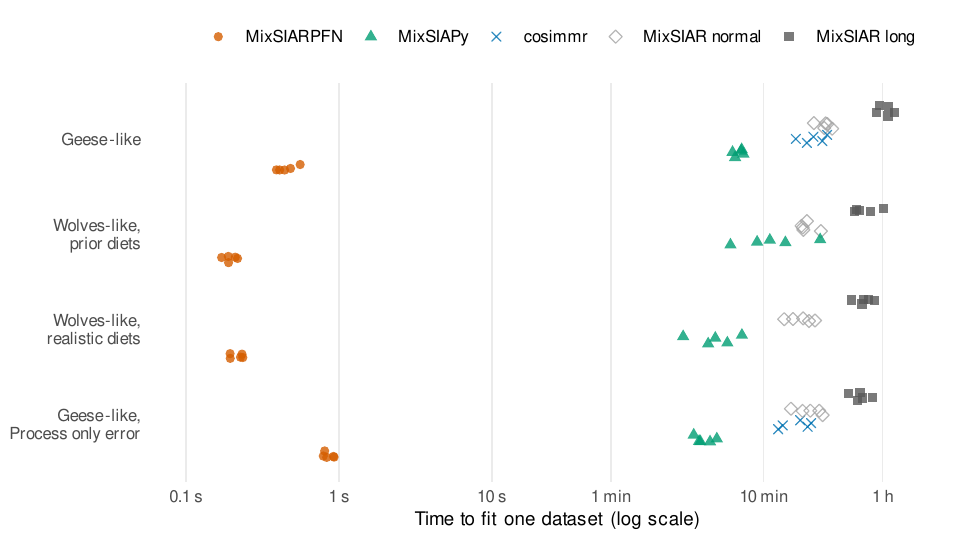}
\caption{Time to fit one dataset, for five held-out datasets of each test design, on an otherwise idle machine (log scale). MixSIARPFN ran through its R package on one processor core, after a first fit had loaded the network; MixSIAPy used three cores and the other methods one, with MixSIAR's three chains run one after another. cosimmr cannot fit the wolves-like designs. MixSIARPFN fitted each dataset in under a second, three to four orders of magnitude faster than MixSIAR and two to three faster than the other fast methods.}
\label{fig:speed}
\end{figure}

Through the R package on one processor core, a MixSIARPFN fit took a median of 0.20 to 0.76 seconds, depending on the design (Fig.~\ref{fig:speed}). The first fit in each R session takes 4.2 to 4.8 seconds, because it starts Python and loads a network (Intel Xeon Gold 6542Y, otherwise idle). MixSIAR, running its three chains one after another on one core, took 18 to 25 minutes at its normal run length and 42 to 60 minutes at its long one, so MixSIARPFN was 1,600 to 5,400 and 3,300 to 14,100 times faster. Running MixSIAR's chains in parallel on three cores would reduce these factors at most threefold. MixSIAPy took 4 to 10 minutes on three cores and cosimmr 17 to 20 minutes on one, so MixSIARPFN was 340 to 3,000 and 1,300 to 2,900 times faster than these.

\subsection{An illustration of the advantages of MixSIARPFN}

We performed a full sensitivity analysis of MixSIAR's alligator dataset, refitting it 3,600 times with MixSIARPFN under every combination of model structure, error structure, discrimination factors and prior, and observed the effect on the dietary proportions. The fits took 62 minutes in all (0.8 seconds each). With MixSIAR, at the median time of a converged run for these models (77 minutes), they would have taken about 190 days.

\begin{figure}[t]
\centering 
\includegraphics[width=\textwidth]{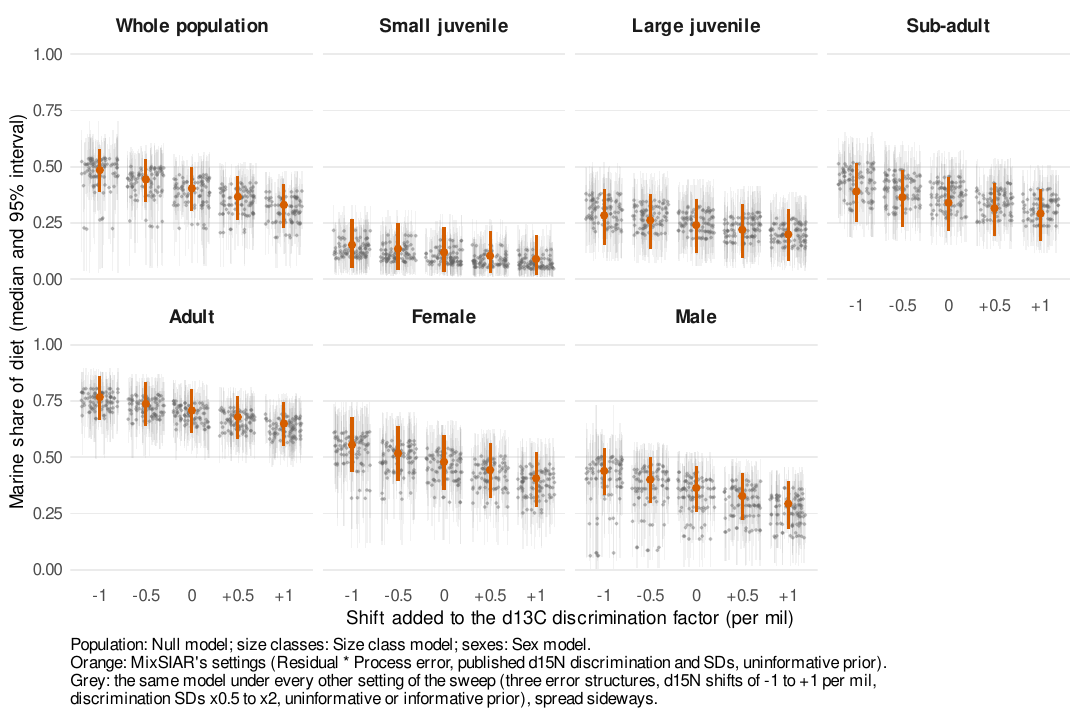}
\caption{A sensitivity analysis of MixSIAR's alligator example that is impractical with Markov chain Monte Carlo: the marine share of the diet, by size class and sex and for the whole population, from 3,600 MixSIARPFN fits over model structures, error structures, discrimination factors and priors. Orange: posterior median and 95\% interval with MixSIAR's settings, against the shift added to the $\delta^{13}$C discrimination factor; grey: every other setting of the sweep at that shift. The ordering of size classes, and the larger marine share of adults, held in every fit, and the population's marine share depended mostly on the $\delta^{13}$C discrimination factor.}
\label{fig:sweep}
\end{figure}

The study's main conclusions were robust (Fig.~\ref{fig:sweep}). In every fit, adults ate more marine prey than small juveniles, the marine share rose with every size class, and alligators caught in marine habitat ate more marine prey than those caught in freshwater. Freshwater prey made up most of the population's diet in 88\% of fits, and the marine share rose clearly with length in 89\%.

The population's marine share depended most on the $\delta^{13}$C discrimination factor, moving by about 15 percentage points between shifts of $-1$ and $+1$\textperthousand, more than under any other change. Females ate more marine prey than males in the model with sex alone, but less in the model with sex and size class together, because 67 of the 76 juveniles were male. No fit separated the sexes clearly. Changing the discrimination standard deviations, which are small compared with the sources' own variation, made almost no difference.

\section{Discussion}

A single pass through a network trained once on simulations from MixSIAR's model gave posteriors 3\% to 4\% less accurate than MCMC on held-out data with MixSIAR's usual error structures, with well-calibrated, slightly conservative intervals, in under a second on one processor core. This makes routine what was impractical: fitting every candidate model, analysing hundreds of populations or sites, and the sensitivity analyses over discrimination factors and priors that the literature recommends \citep{bond2011sensitive, phillips2014best, swan2019evaluating}. The alligator sweep took an hour rather than half a year. There is also no chain to check, whereas MixSIAR's standard runs failed the pre-registered convergence check on 17\% to 66\% of our test datasets.

MixSIAPy is a new, efficient sampler of MixSIAR's exact model. We found it to be as accurate as converged MixSIAR, it converged far more often, and it is the method of choice when a full MCMC run is required. MixSIARPFN gives up 3\% to 4\% of accuracy, partly as wider intervals, for answers in under a second, so a natural workflow uses both: MixSIARPFN to explore models and data, and a sampler for the final model. cosimmr was neither fast nor accurate here, because it fits a simpler model than the one that generated the data.

The approximation is weakest where data are unlike the training simulations. With heavy-tailed or skewed errors on the geese-like design, its gap to MCMC grew from 4\% to 13\% and 11\%. On the geese and wolves examples, MixSIARPFN differed more from MixSIAR than elsewhere, but MixSIAR's model describes these data poorly and its own answer moved as much under a different error structure. In such cases MCMC returns the exact posterior of a model that does not fit, while the network extrapolates beyond its training data, and neither answer should be trusted until the model is improved.

There are other limitations. The Process only network was 8\% to 12\% behind MCMC where MCMC could be trusted, and is the first candidate for further training. The networks give marginal posteriors only, so the analyses in Table~\ref{tab:features} that need the joint posterior still need MCMC (our joint-diet training is a first step). Finally, the package rejects designs outside the training envelope, but does not yet check whether values within a design, such as discrimination standard deviations or consumers outside the mixing polygon, are unlike the training data.

Some limits of the envelope could become warnings instead. In a test of the Residual~$\times$~Process network beyond its envelope (Appendix S10), it stayed calibrated with up to 5,000 consumers, 6 tracers or 10 sources, and with 5 or 6 tracers or 9 or 10 sources it was as accurate as within the envelope. With more than 1,000 consumers, however, its posteriors were up to 30\% wider than MCMC's and gained little from the extra consumers. The limits of 20 levels per factor and two factors are fixed by the architecture.

We therefore suggest using MixSIARPFN as MixSIAR would be used, but more freely, and checking the final model with MCMC (MixSIAR, or MixSIAPy for speed) where our results show the approximation is weakest or the joint posterior is needed. These cases are consumers near or outside the mixing polygon \citep{smith2013polygons}, sources the data cannot distinguish \citep{fry2013underdetermined}, posteriors close to zero, Process only error, and conclusions resting on differences between groups or sums of sources. Because MixSIAR's likelihood can be evaluated, importance sampling from the network's posterior could flag the fits that need MCMC, and we plan to add this to the package.

More broadly, amortised inference can serve a whole, widely used family of ecological models through one set of networks reused by every analysis, as \citet{kipnis2026metabeta} showed for generalised linear mixed models. Its cost is front-loaded: training the released networks used 112~kWh, several times the energy of all the MCMC in this study (Appendix S7), and this is repaid only as analyses reuse them. What remains is to close the gap to exact inference on data unlike the training simulations, so that a fast answer can always be trusted or flagged.

\section*{Use of artificial intelligence tools}

We used Claude Code (Anthropic), with the Claude Opus 5.5 model, as a programming and writing assistant throughout this work: to write and test the simulation, training and package code, run and score the analyses, search and verify the literature, and draft the manuscript. All code, analyses and text were directed, reviewed and checked by the authors, who take full responsibility for them. Commits made with the assistant are marked as such in the repositories' version histories.

\section*{Acknowledgements}

We thank the developers and users of MixSIAR, whose example data and analyses made this comparison possible, and Donald L. Phillips for his contributions to MixSIAR and to the mixing-model methods it builds on.

\section*{Funding}

Andrew Parnell's work was supported by: the UCD-Met \'Eireann Research Professorship Programme (28-UCDNWPAI); a Research Ireland, Northern Ireland's Department of Agriculture, Environment and Rural Affairs (DAERA), UK Research and Innovation (UKRI) via the International Science Partnerships Fund (ISPF) under Grant number [22/CC/11103] at the Co-Centre for Climate + Biodiversity + Water; Decarb AI, supported by AIB and Research Ireland: an Innovate for Ireland Centre (25/I4I-TC/13542); and Taighde \'Eireann -- Research Ireland, under Grant number [23/RC/13506] at the Research Ireland Centre - Rinn Artificial Intelligence.

\section*{Data availability statement}

The MixSIARPFN R package is available at \url{https://github.com/andrewcparnell/MixSIARPFN}, with documentation at \url{https://andrewcparnell.github.io/MixSIARPFN/}. The three trained networks are archived on Zenodo (\url{https://doi.org/10.5281/zenodo.24002247}). The code, simulated datasets, results and manuscript sources for this paper, including the held-out test protocol and the code that simulates the training data and trains the networks, are available at \url{https://github.com/andrewcparnell/MixSIARPFN-paper}.

\bibliographystyle{apacite}
\bibliography{references}

\clearpage
\setcounter{figure}{0}\setcounter{table}{0}
\renewcommand{\thetable}{S\arabic{table}}
\renewcommand{\thefigure}{S\arabic{figure}}
\begin{center}{\Large\bfseries Supporting Information}\end{center}

\section*{Appendix S1. The model, the supported envelope and the training designs}
\addcontentsline{toc}{section}{Appendix S1. The model, the supported envelope and the training designs}

\textbf{The model.} For consumer $i = 1,\dots,N$ with tracer values $\mathbf{y}_i = (y_{i1},\dots,y_{iJ})$ and $K$ sources, MixSIAR writes the diet proportions as 
\[
\mathbf{p}_i = \mathrm{ilr}^{-1}(\boldsymbol{\eta}_i), \qquad
\boldsymbol{\eta}_i = \boldsymbol{\beta}_0 + \boldsymbol{\beta}^{(1)}_{f_1(i)} + \boldsymbol{\beta}^{(2)}_{f_2(i)} + \boldsymbol{\gamma}\, x_i ,
\]
where $\mathrm{ilr}$ is the isometric log-ratio transform to $\mathbb{R}^{K-1}$ \citep{egozcue2003ilr}. Here $\boldsymbol{\beta}_0$ is the global intercept, $f_1(i)$ and $f_2(i)$ are the levels of up to two factors, and $x_i$ is an optional continuous covariate. A fixed factor's first level has effect zero. Its other levels have independent $\mathrm{N}(0, 1)$ priors on the log-ratio scale. A random factor has $\beta^{(r)}_{\ell k} \sim \mathrm{N}(0, \sigma_r^2)$ with $\sigma_r \sim \mathrm{U}(0, 20)$. The second factor may be crossed with the first or nested within it. The prior on $\mathbf{p}$ at the global level is Dirichlet with weights $\boldsymbol{\alpha}$, which MixSIAR induces on $\boldsymbol{\beta}_0$.

With source means $\mu_{kj}$ and variances $\omega^2_{kj}$ (from the source data, by level of one factor when sources are given by a factor), discrimination means $\lambda_{kj}$ and variances $\tau^2_{kj}$, and elemental concentrations $c_{kj}$ (all equal to 1 without concentration dependence), the mean tracer value is 
\[
\mathrm{E}(y_{ij}) = \frac{\sum_k p_{ik}\, c_{kj}\,(\mu_{kj} + \lambda_{kj})}{\sum_k p_{ik}\, c_{kj}} ,
\]
and $\mathbf{y}_i$ is multivariate normal. Its covariance is given by one of three error structures \citep{stock2016error}. Process only uses the source and discrimination variances weighted by the squared proportions, $\sum_k p_{ik}^2(\omega_{kj}^2 + \tau_{kj}^2)$ for tracer $j$, with no covariance between tracers. (MixSIAR's JAGS code squares the proportions themselves, not the concentration-weighted ones.) Residual only uses a single residual covariance matrix with a Wishart prior (a gamma prior on the precision for one tracer). Residual~$\times$~Process multiplies the process variance of tracer $j$ by $\xi_j \sim \mathrm{U}(0, 20)$ (\texttt{resid.prop}). Source means and variances are estimated from the source data within the same model, with MixSIAR's vague priors \citep{stock2018mixsiar}. MixSIARPFN first summarises raw source samples to means, standard deviations and sample sizes.

\textbf{The supported envelope.} Table~\ref{tab:envelope} gives the designs MixSIARPFN accepts. These are the designs the networks were trained on. Every MixSIAR isotope example falls inside the envelope. Data outside it are refused with a message naming the limit exceeded (Appendix~S10 tests the networks beyond it).

\begin{table}[h]
\centering\small 
\caption{The supported envelope: the designs the networks were trained on and the package accepts. Every MixSIAR isotope example falls inside it.}
\label{tab:envelope}
\input{tables/si_envelope.tex}
\end{table}

\textbf{Training designs.} Each training dataset was drawn in three stages: the design, the sources, and then the parameters and consumers. The full configuration is in the repository (\texttt{configs/general.yaml}). The number of sources was uniform on 2 to 8, and the number of tracers on 1 to 4. The effects had relative weights of 1 for none, one fixed and one random. They had weights of 0.7 for each of two fixed, fixed and random, two random crossed, and two random nested. The numbers of levels were drawn within the envelope.

The total number of consumers was log-uniform on 1 to 1,000, split unevenly over levels. A continuous covariate was included with probability 0.35. Sources were given by a factor with probability 0.3. Concentration dependence was included with probability 0.3, with concentration ratios up to 100.

The source means were spread and overlapping over the range seen in MixSIAR's examples. Source standard deviations varied by a factor of about 1.5 between sources. Discrimination factors had standard deviations from 0 to 0.6. Source sample sizes ran from 2 to 300.

The Dirichlet prior weights were all 1 (probability 0.2), all equal to one value from 0.05 to 20 (0.3), or independent from 0.005 to 100 (0.5). This variety meant the networks learnt to read the prior from the data table. Finally, the parameters were drawn from MixSIAR's priors given the design, and the consumers from the model.

\section*{Appendix S2. The networks and their training}
\addcontentsline{toc}{section}{Appendix S2. The networks and their training}

\textbf{Architecture.} Each network embeds every cell of the data table as a vector of 384 numbers. It passes the table through eight layers, each with six attention heads and a feed-forward block of width 768. Attention alternates between the columns of a row and the rows of a column, as in TabPFN \citep{hollmann2025accurate}. Query rows attend to the data rows but not to each other. Each answer therefore depends only on the data and its own query.

Factor levels are embedded by vectors drawn at random from a pool of 20 for each dataset. The network therefore cannot attach meaning to a level's label. The package averages each answer over eight such draws. The output for a diet query is a probability for each of 1,000 bins: 500 equally wide on $[0, 1]$ and 500 equally wide on the logit scale between $-10$ and 10. Standard deviations and \texttt{resid.prop} both lie on $[0, 20]$ under MixSIAR's priors, and use 400 even and 200 logarithmic bins. Covariate slopes use 1,000 bins, finest near zero.

\textbf{Invariance.} Rows carry no positional information. By construction, a network's output therefore does not depend on the order of the consumers or of the sources. We checked the source order numerically by refitting development datasets with their sources in random orders. Posterior summaries changed by less than $10^{-6}$. Averaging over 32 or 128 level assignments instead of eight changed CRPS by less than 0.5\% (Fig.~\ref{fig:ablations}).

\textbf{Training.} Each step used 131,072 table cells (a batch of datasets of mixed sizes). We used schedule-free AdamW \citep{defazio2024road} at learning rate $3\times10^{-4}$ after 2,000 warm-up steps, with gradient clipping at 1 and bfloat16 arithmetic. The loss was the negative log score of the true value under the predicted histogram. It was averaged within each kind of query (global diets, level diets, combinations, covariate values, standard deviations, slopes and residual terms) and then across kinds. This stopped the many level queries from dominating.

\begin{figure}[h]
\centering 
\includegraphics[width=\textwidth]{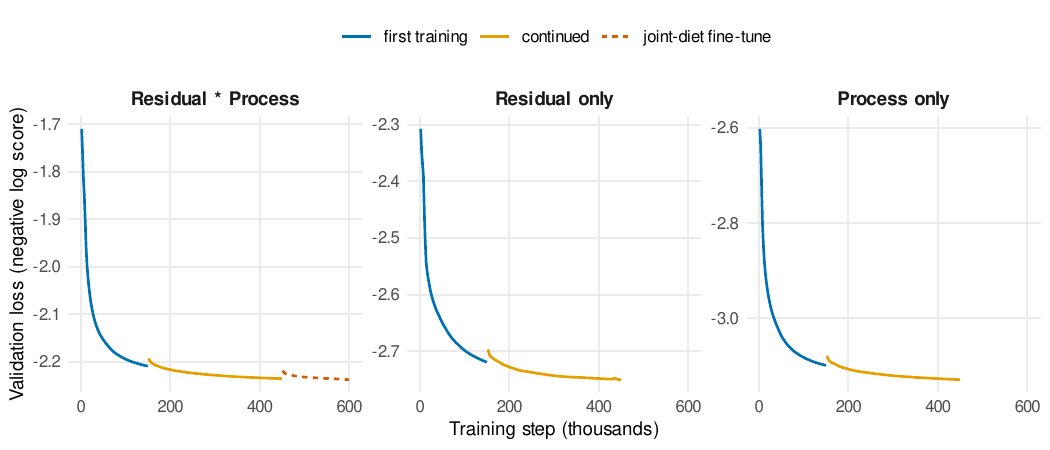}
\caption{Validation loss (negative log score on 24,000 fixed simulated datasets; lower is better) through training of the three released networks. Losses were still falling slowly when training stopped, consistent with the finding that training length was the largest influence on accuracy (Fig.~\ref{fig:ablations}). The joint-diet fine-tune of the Residual~$\times$~Process network also lowered the loss of its ordinary marginal queries.}
\label{fig:training}
\end{figure}

Validation every 2,000 steps used a fixed set of 24,000 datasets (400 in each of 60 strata of sources, tracers, effects, covariate and consumers). These were simulated once with a separate seed and never used in training. Training ran in stages (Table~\ref{tab:training}, Fig.~\ref{fig:training}). The released checkpoint of each network is the one with the lowest validation loss.

\begin{table}[h]
\centering\small 
\caption{Training of the released networks. Steps are those of each stage; the best validation loss is the negative log score on the fixed validation set (lower is better), which kept improving to the end of each stage. Validation loss is not comparable between networks, whose parameters differ.}
\label{tab:training}
\resizebox{\textwidth}{!}{\input{tables/si_training.tex}}
\end{table}

\textbf{Calibration on the validation set.} On the validation set, the released networks' central 50\%, 90\% and 95\% intervals contained the true values at their nominal rates (Table~\ref{tab:calibration}). This held overall and in every subgroup of sources, tracers, effects, covariate and consumers. All three networks are calibrated to within 0.2 percentage points. This is a check on average over the training distribution. The simulation-based calibration below checks each kind of quantity separately.

\begin{table}[h]
\centering\small 
\caption{Calibration of the released networks on the validation set: the share of true values inside the central 50\%, 90\% and 95\% intervals, the Kolmogorov--Smirnov distance of the probability integral transform from uniform (0 for perfect calibration), and the lowest and highest 95\% coverage across 14 subgroups by number of sources, tracers, effects, covariates and consumers. All three networks are calibrated to within 0.2 percentage points.}
\label{tab:calibration}
\resizebox{\textwidth}{!}{\input{tables/si_calibration.tex}}
\end{table}

\textbf{Simulation-based calibration across the envelope.} The validation set checks calibration on average. Simulation-based calibration checks it for each kind of quantity \citep{talts2018validating, modrak2025simulation}. For each network we drew 2,000 datasets from the training distribution (400 designs, five datasets each; seed 5001). We fitted each with the released network as the package does, averaging over eight level assignments. We then computed the probability integral transform (PIT) of every true value in its marginal posterior. The PIT is exact, from the histogram's distribution function.

A true value in the first or last bin was given a PIT drawn uniformly within that bin's probability. These bins are within $4.5\times10^{-5}$ of 0 or 1 for proportions, and below 0.001 for standard deviations. The prior on random-effect standard deviations pushes many level diets to within $10^{-10}$ of 0. Without this rule, those diets would pile up at a PIT of 0.

\begin{figure}[h]
\centering 
\includegraphics[width=\textwidth]{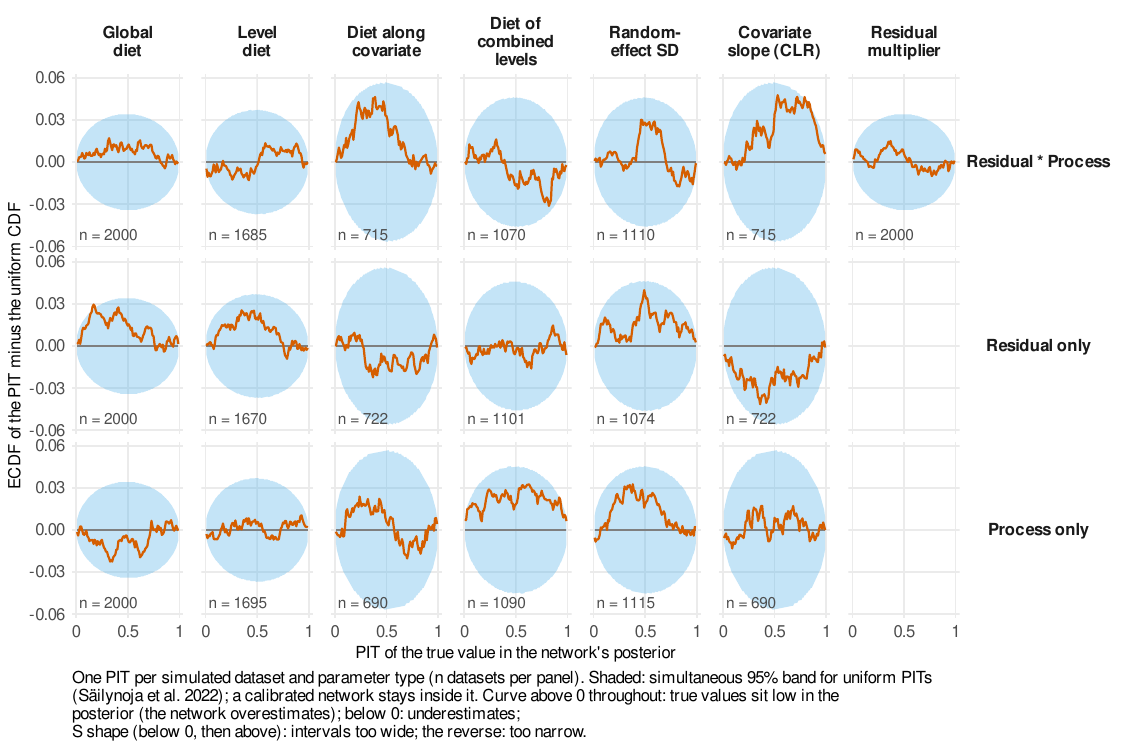}
\caption{Simulation-based calibration: the empirical distribution function of the PIT of the true value in each network's posterior, minus the uniform distribution function, for each network (rows) and kind of quantity (columns), with one PIT per dataset; shaded, the simultaneous 95\% band for uniform PITs. A curve above zero means true values sit low in the posterior (the network overestimates). The networks stay inside or close to the bands everywhere.}
\label{fig:sbc}
\end{figure}

Table~\ref{tab:sbc} summarises the PITs. Fig.~\ref{fig:sbc} compares the distribution of one PIT per dataset and quantity with the simultaneous 95\% band for uniformity \citep{sailynoja2022graphical}. Coverage was within 0.6 percentage points of nominal everywhere. Three of 19 combinations left the band, by small margins. For the Residual~$\times$~Process network these were covariate slopes (largest deviation 0.048) and level diets (0.014). For the Residual only network it was global diets (0.03), where true values sit slightly low in the posterior. The Residual only network also slightly overestimates random-effect standard deviations (mean PIT 0.48).

\begin{table}[h]
\centering\small 
\caption{Simulation-based calibration of the released networks on 2,000 datasets each from the training distribution. KS distance: Kolmogorov--Smirnov distance of all PITs from uniform; 50\%, 90\%, 95\%: coverage of central intervals; Mean PIT: 0.5 for an unbiased network; Inside band: whether the distribution of one PIT per dataset stays inside the simultaneous 95\% band for uniformity. All coverages are within 0.6 points of nominal.}
\label{tab:sbc}
\resizebox{\textwidth}{!}{\input{tables/si_sbc.tex}}
\end{table}

\section*{Appendix S3. The held-out test protocol and leakage audit}
\addcontentsline{toc}{section}{Appendix S3. The held-out test protocol and leakage audit}

We chose between network versions (v1, v2, a fine-tune on common designs called q3, and v3) using development simulation sets. These were 450 datasets of the geese-like and wolves-like designs, in two batches. Results on those sets are therefore optimistic for the version chosen. Before any test dataset was generated, we wrote and committed a protocol.

The protocol fixed the following:
\begin{itemize} 
\item the four test designs and their seeds (3001 to 3004; development seeds were different);
\item 100 datasets per design, selected by the same nearest-to-the-real-data rule as in development;
\item the methods and their settings;
\item the metrics, the boundary rule for coverage, and the paired bootstrap (4,000 resamples, seed 1);
\item the primary comparison (converged long MixSIAR runs) and the secondary one (MixSIAR normal runs, all datasets);
\item that every method would be scored once.
\end{itemize} 
We made one change to the analysis after scoring. The protocol specified a classical paired bootstrap for intervals. We report the Bayesian bootstrap \citep{rubin1981bayesian} instead, which gives credible intervals and suits a Bayesian analysis. No point estimate changed, and no interval end moved by more than 0.5 percentage points. Two dated addenda were also committed before scoring. They fixed the settings of MixSIAPy and cosimmr and the design of the timing benchmark. The protocol and its addenda are in the repository (\texttt{data/processed/splits/protocol.md}).

The leakage audit (Table~\ref{tab:leakage}) was computed from the files before any test-set score existed. All seven checks passed. The networks are trained only on datasets simulated from the prior during training. No test dataset can therefore have been seen in training. The remaining risk is selection on the test sets, which the protocol rules out.

\begin{table}[h]
\centering\small 
\caption{Leakage audit, run before scoring the held-out test sets. All seven checks passed.}
\label{tab:leakage}
\input{tables/si_leakage.tex}
\end{table}

\section*{Appendix S4. Convergence, development results and the 42 benchmark designs}
\addcontentsline{toc}{section}{Appendix S4. Convergence, development results and the 42 benchmark designs}

\textbf{Convergence.} The pre-registered rule took a MixSIAR run as converged when its largest Gelman--Rubin diagnostic over all parameters was at most 1.1. By that rule MixSIAR often failed on the test sets (Table~\ref{tab:convergence}). Failures were most common on the wolves-like designs, where many pack diets are close to a single source. In response to reviewers we also applied the current recommendation \citep{vehtari2021rank}. This requires rank-normalised split $\hat{R}$ below 1.01 and bulk and tail effective sample sizes above 400 for every diet proportion.

MixSIAR's runs keep 1,000 thinned draws per chain, and almost never pass this strict rule. MixSIAPy's no-U-turn sampler passed on 89\% to 100\% of the datasets of the three main designs. It passed on 63\% of the Process only design. This is why the main text adds MixSIAPy as a reference on all datasets. On datasets where MixSIAPy passed the strict rule and MixSIAR long passed the pre-registered one, their mean CRPS agreed to within 0.7\% on every design. The two samplers therefore target the same posterior.

\begin{table}[h]
\centering\small 
\caption{Share of the 100 test datasets per design on which each MCMC method passed a convergence check. Pre-registered: largest Gelman--Rubin diagnostic at most 1.1 (MixSIAR, over all parameters) or largest rank-normalised $\hat{R}$ over the diet proportions at most 1.1 (MixSIAPy). Strict: rank-normalised split $\hat{R}$ below 1.01 and bulk and tail effective sample sizes above 400 for every diet proportion. MixSIAR's runs rarely pass the strict rule; MixSIAPy passes it on most datasets except for Process only error.}
\label{tab:convergence}
\resizebox{\textwidth}{!}{\input{tables/si_convergence.tex}}
\end{table}

\textbf{Which datasets converged.} Whether MixSIAR converges depends on the dataset, so the converged datasets are not a random sample of each design (Table~\ref{tab:selection}). On the wolves-like designs, the datasets where MixSIAR long failed to converge were, if anything, easier for MixSIARPFN (lower CRPS). On the Process only design they were harder. On both subsets of the first three designs, MixSIAR's normal runs were 0\% to 4\% more accurate than MixSIARPFN. On the Process only design, they were 10\% more accurate where long runs converged. They were 18\% less accurate where long runs did not converge. This is why MixSIARPFN is ahead of MixSIAR's normal runs over all 100 Process only datasets. This is why the main text reports both comparisons.

\begin{table}[h]
\centering\small 
\caption{Held-out test datasets split by whether MixSIAR's long run converged. CRPS is MixSIARPFN's mean continuous ranked probability score over all diet proportions (lower is better); the last two columns give the CRPS of MixSIAR's normal runs relative to MixSIARPFN's (negative: MixSIAR normal more accurate). The converged and unconverged datasets differ, so comparisons restricted to converged runs are not representative of a whole design.}
\label{tab:selection}
\resizebox{\textwidth}{!}{\input{tables/si_selection.tex}}
\end{table}

\textbf{Further metrics on the test sets.} These post hoc metrics are computed by the script \texttt{19\_reanalysis.R} in the repository. The root mean squared error of MixSIARPFN's posterior means was 2\% to 3\% above MixSIAPy's on the three main designs. It was below MixSIAPy's for Process only (0.115 against 0.122). MixSIARPFN's posterior means differed from MixSIAPy's by a median of 0.9 to 1.4 percentage points. Of these differences, 90\% were below about 5 points and 2\% to 3\% exceeded 10 points. The largest were 25 to 48 points.

\textbf{Coverage near the boundary.} In the wolves-like design with prior diets, 23\% of true proportions lie within $4.6\times10^{-5}$ of 0 or 1. This is beyond the networks' finest histogram bin, which ends at a logit of $\pm10$ ($4.5\times10^{-5}$). The pre-registered rule counts such values as on the boundary. With this rule, MixSIARPFN's coverage there is 97\%. Without it, coverage is 79\%, against 96\% for the samplers. On the other designs the rule makes no difference. Fig.~2b of the main text shows coverage intervals from a Bayesian bootstrap over datasets. Table~\ref{tab:fast} summarises every method compared on the test sets, against converged MixSIAR.

\begin{table}[h]
\centering\small 
\caption{The methods compared on the held-out test designs. CRPS: percentage difference in mean continuous ranked probability score from converged MixSIAR long runs on the datasets where those converged (ranges across the designs each method can fit). Coverage: share of 95\% intervals containing the true proportion. Time per fit: medians on an idle machine; MixSIARPFN through its R package on one core, after a first fit (4.2 to 4.8 s) that loads the network. MixSIARPFN was the only method under a second; MixSIAPy was as accurate as converged MixSIAR but took minutes.}
\label{tab:fast}
\resizebox{\textwidth}{!}{\input{tables/table2_fast.tex}}
\end{table}

\textbf{Development results by network version.} Table~\ref{tab:development} gives the gap between each network version and converged MixSIAR on the development sets. Longer training gave the largest improvement. These numbers guided the choice of v3, so they are optimistic for it. On the held-out wolves-like set with prior diets, the gap was 4.3\%, against 1.9\% in development.

\begin{table}[h]
\centering\small 
\caption{Development results: the CRPS of each network version relative to converged long MixSIAR runs (\%; positive means less accurate than MixSIAR), on the development datasets where MixSIAR converged, both batches pooled. v1: first training (150,000 steps); v2: continued to 450,000 steps; q3: v2 fine-tuned on common designs; v3: the released networks (v2 with the joint-diet fine-tune of the Residual~$\times$~Process network). Longer training gave the largest improvement. These sets guided the choice of v3, so its numbers are optimistic.}
\label{tab:development}
\input{tables/si_development.tex}
\end{table}

\textbf{The benchmark designs.} Table~\ref{tab:benchmark} lists the designs used to test the networks across the supported envelope. For each, it gives the Wasserstein-1 distance between the MixSIARPFN and converged MixSIAR posteriors, and each method's coverage of the true diets. Each design is a single simulated dataset. The table is therefore illustrative rather than a sample of datasets per design. These designs were also used during development to compare network versions, so they are not held out.

Two families reuse the sources of MixSIAR's geese and killer whale examples with other designs. The general family draws new sources. Of the 47 designs simulated, seven are excluded. In six, MixSIAR did not converge: two had a largest Gelman--Rubin diagnostic of 1.29, three failed even at extreme run length, and one extreme run had not finished at the time of analysis. In the seventh, MixSIAR could not fit the model because its inverse log-ratio transform overflows.

\input{tables/si_benchmark_long.tex}

\section*{Appendix S5. MixSIAR's examples and posterior predictive checks}
\addcontentsline{toc}{section}{Appendix S5. MixSIAR's examples and posterior predictive checks}

\textbf{The examples, model by model.} We fitted MixSIARPFN to every model of MixSIAR's nine isotope examples, with each example's settings (Table~\ref{tab:examples}). We compared it with a converged MixSIAR run on the same data. We used long runs, or very long or extreme runs where long ones did not converge. For the four examples with raw source data, both methods used the same source summaries. MixSIARPFN summarises raw samples to means, standard deviations and sample sizes. MixSIAR's raw-data model estimates the source distributions jointly with the diets, so it is a different model. For Palmyra, MixSIAR's own fits with raw and summarised source data differ by 0.096 in mean absolute difference of posterior means. That is more than MixSIARPFN differs from MixSIAR (0.031). The geese and wolves examples differed most.

\begin{table}[h]
\centering\small 
\caption{MixSIARPFN against converged MixSIAR on each model of MixSIAR's isotope examples. Mean abs. diff.: mean absolute difference between the two methods' posterior means over all diet proportions; W1: mean Wasserstein-1 distance between the posteriors; SD ratio: median ratio of MixSIARPFN's posterior standard deviation to MixSIAR's; PFN (s): MixSIARPFN's fitting time in seconds; MixSIAR (min): MixSIAR's run time in minutes. The two methods agree closely on every model except geese and wolves, at a small fraction of the time.}
\label{tab:examples}
\resizebox{\textwidth}{!}{\input{tables/si_examples.tex}}
\end{table}

\textbf{Posterior predictive checks.} We asked whether MixSIAR's model describes the geese, wolves and killer whale data. To do so, we reran MixSIAR's own JAGS model on its saved data with one added line. This line draws replicated consumers from the likelihood at every iteration (3,000 draws, MixSIAR's ``very long'' setting). For each group (geese) or pack (wolves), we computed each tracer's mean and standard deviation and the correlation between the two tracers. We compared these statistics in the real data with their distribution over the replicated datasets. This gave 40 statistics for geese and wolves, and 5 for the single-group killer whale data. We also computed a posterior predictive $p$-value for the total log density. As controls, we ran the same check on six datasets simulated from the model itself for each of the geese-like and wolves-like designs.

\begin{table}[h]
\centering\small 
\caption{Posterior predictive checks of MixSIAR's fitted model. Statistics: the number of group summaries checked (tracer means, standard deviations and between-tracer correlations per group); Outside 95\% (99\%): how many fell outside the central 95\% (99\%) range of replicated datasets; Log density $p$: posterior predictive $p$-value of the total log density. Data simulated from the model rarely had more than two extreme statistics, but the real geese and wolves data had 11 and 5: MixSIAR's model does not reproduce them.}
\label{tab:ppc}
\input{tables/si_ppc.tex}
\end{table}

Data simulated from the model rarely had more than two extreme statistics (Table~\ref{tab:ppc}). The real geese and wolves data had 11 and 5. MixSIAR's model therefore does not reproduce them. The whole-data log density was unremarkable in every case, so a global check would not catch these misfits. They are in the group means and spreads. The networks were trained only on data the model could have generated. Where the model misfits, the exact posterior and an amortised approximation to it can therefore differ more.

\textbf{Why MixSIARPFN differs from MixSIAR on the geese and wolves examples.} We examined the largest differences with the script \texttt{23\_example\_gaps.R}. On geese, they are in the split of the diet between \emph{Enteromorpha} and \emph{Ulva lactuca}, two sources close in isotope space. They occur in groups 2, 3, 5 and 6, which have 20 to 74 birds each. For group 5, MixSIAR's posterior mean for \emph{U.\ lactuca} was 0.50 (standard deviation 0.20) and MixSIARPFN's was 0.23 (0.24). MixSIAR's posteriors for these proportions are wide (median standard deviation 0.15, against 0.03 for the other proportions). They are often bimodal, and put up to 18\% of their draws below 0.01. The two methods agree on the sum of the two sources to within 0.036 in every group. On wolves, the largest difference is for salmon in pack 7, which has four wolves (0.14 against 0.36).

These differences are not MCMC noise. Each MixSIAR chain alone agreed with MixSIAR's reference fit to within 0.025 on every proportion of both examples. So did a second MixSIAR run with another seed, MixSIAR's runs of other lengths, and MixSIAPy. Nor did anything on MixSIARPFN's side reduce the differences. Averaging over 32 or 128 level assignments instead of 8, or other random seeds, moved any proportion by at most 0.03.

The error model matters instead. We refitted the geese data by MCMC with Residual~$\times$~Process error instead of Residual only. This moved MCMC's answer as far as MixSIARPFN's difference (mean 0.074, largest 0.25). MixSIARPFN's answer is close to that alternative (mean 0.025, largest 0.093). Under Residual~$\times$~Process error, MixSIARPFN agreed with MCMC fitted to the same model (94\% of proportions within 5 points). That MCMC fit was, however, not fully converged ($\hat{R}$ 1.02).

The split between these two sources therefore depends on the error model as much as on the method. On such data, users should interpret the sum of the similar sources, or confirm the split with MCMC. We also measured the share of proportions within 5 points of MixSIAR. By this measure the real geese data were worse than all 100 simulated geese-like test datasets. The real wolves data were in the worst 10\% of the wolves-like ones.

\section*{Appendix S6. Joint-diet training}
\addcontentsline{toc}{section}{Appendix S6. Joint-diet training}

Marginal posteriors do not give the posterior of a sum or difference of proportions, or their correlations. To obtain joint draws, we trained the Residual~$\times$~Process network for a further 150,000 steps with an additional query type. This query gives the posterior of one diet's proportions one source at a time. Each proportion is a share of what the previously drawn sources leave (stick-breaking), conditioned on the shares already drawn. Six extra input features carry the conditioning values. In this stage 40\% of diet queries remained unconditional, so the network kept answering ordinary marginal queries. Draws are made by sampling the sources in a random order, one conditional query at a time.

A Residual only network was trained with the same additional task. It was stopped early, after 90,000 steps, when its validation loss on ordinary marginal queries stopped improving. The fine-tune also added 150,000 training steps, and longer training was the largest influence on accuracy (Fig.~\ref{fig:ablations}). The gain in the marginals therefore cannot be attributed to the joint task itself. We did not train a control with the same extra steps and no joint task.

\begin{table}[h]
\centering\small 
\caption{Joint draws against converged MixSIAR on the development sets (both batches pooled; diets are all global and group diets). Mean abs. diff. in correlations: the mean absolute difference between the pairwise posterior correlations from the networks' joint draws and from MixSIAR's draws. Strongest pair $r$: the mean correlation of each diet's most correlated pair of sources. Coverage of sum: the share of true values of sums of two proportions inside the 95\% interval. The Residual~$\times$~Process network (wolves-like rows, released) reproduced MixSIAR's correlations closely; the Residual only network (geese-like row, not released) produced correlations of the right sign but too weak.}
\label{tab:joint}
\resizebox{\textwidth}{!}{\input{tables/si_joint.tex}}
\end{table}

On the wolves-like development sets, the joint-trained Residual~$\times$~Process network reproduced MixSIAR's posterior correlations closely (Table~\ref{tab:joint}). It also gave calibrated intervals for sums of proportions. Its ordinary marginal queries became 1.0\% to 1.5\% more accurate (paired against v2). Marginals computed from joint draws became 1.7\% to 2.1\% more accurate. The gain was largest (3\% to 4\%) for proportions whose posterior correlation with another source exceeded 0.85 in absolute value. The package uses the ordinary marginals, which take about a second. Joint draws took about two minutes for 3,000 draws of 12 diets.

On the geese-like design, the Residual only network's conditional predictions were too weak. The strongest pairwise correlation averaged $-0.67$, against MixSIAR's $-0.84$. Its marginals also became less accurate than v2's, so it was not released. A diagnosis on further simulated sets ruled out the number of sources and concentration dependence as main causes. Even the first conditional step responded about half as strongly to the previously drawn source as MixSIAR's draws imply (a correlation of 0.31 against 0.56). The same design with Residual~$\times$~Process error showed a similar but smaller weakness. The geese design has many consumers per group (31 on average), so each group diet is tightly determined and its conditional structure fine. We expect more training on such designs to be needed before joint draws can be offered generally.

\section*{Appendix S7. Compute and energy}
\addcontentsline{toc}{section}{Appendix S7. Compute and energy}

Table~\ref{tab:compute} accounts for the computing used in this study. GPU time is from the training logs. We used NVIDIA L40S GPUs, two per run for the released networks and one for some earlier runs. CPU time for MixSIAR, MixSIAPy and cosimmr is each run's recorded wall-clock time multiplied by its cores. Energy assumes 350~W per GPU (its power limit, so an upper bound) and 250~W per 24-core CPU. Carbon assumes 255~g~CO$_2$ per kWh (the Irish grid in 2023).

Training the released networks used 320 GPU-hours, about 112~kWh. All training, including unreleased attempts, used about 198~kWh. By comparison, the MixSIAR runs of this study used about 3,200 CPU core-hours and 34~kWh. Within this study, therefore, training cost several times more energy than the MCMC it replaced. The saving accrues over the analyses that reuse the networks. Fitting all 400 held-out test datasets with MixSIARPFN took about 2.5 minutes of wall-clock time in total.

\begin{table}[h]
\centering\small 
\caption{Compute and energy. Hours are GPU wall-clock hours for training rows and CPU core-hours otherwise. ``Attempts not released'' are training runs of the same architecture that were superseded (for example, the common-design fine-tune and the larger network); ``Earlier prototypes'' are networks for single example designs that preceded the general networks. Training the released networks is a one-off cost of about 112 kWh; every subsequent fit is nearly free.}
\label{tab:compute}
\input{tables/si_compute.tex}
\end{table}

\section*{Appendix S8. The misspecification study}
\addcontentsline{toc}{section}{Appendix S8. The misspecification study}

The protocol (\texttt{data/processed/splits/protocol\_misspecification.md}) was committed before any of these data were generated. For each design and perturbation, we simulated 200,000 datasets from MixSIAR's prior with a new seed (4001 to 4010). We kept the 20 whose clean consumer summaries were nearest the real data, by the same rule as the test sets. Only then were the selected data perturbed. All 200 datasets were fitted by every method.

Before fitting we checked each perturbation:
\begin{itemize} 
\item the clean selection reproduces the held-out test sets exactly;
\item heavy-tailed deviations follow a $t_3$ distribution (Kolmogorov--Smirnov $p \geq 0.25$);
\item skewed deviations have skewness 0.85, as expected for shape 5;
\item exactly 5\% of consumers (13 geese-like, 3 wolves-like) were moved by 4 standard deviations;
\item discrimination-biased data are shifted by exactly the stated amounts;
\item overstated source precision changes only the reported sample sizes.
\end{itemize} 

MixSIAPy converged (rank-normalised $\hat{R}$ at most 1.01 for every diet proportion) on almost all datasets. The exceptions were 5 of 20 heavy-tailed and 1 of 20 outlier wolves-like datasets. For geese-like data, results on all datasets were the same to within 0.5 points. For wolves-like data, using all datasets changed the heavy-tailed and outlier results to $+2.5\%$ and $-1.5\%$.

We also measured the share of consumers outside the convex hull of the discrimination-corrected source means. Its median was 0.37 to 0.55 for geese-like and 0.66 to 0.77 for wolves-like datasets. It was not related to MixSIARPFN's relative accuracy (Spearman correlations $-0.17$ to $0.26$). Median times per fit were 0.4 to 1.2 seconds for MixSIARPFN on one thread. They were 5 to 6 minutes for MixSIAPy on three cores and 15 to 34 minutes for MixSIAR's normal runs on one core. These times are from a busy machine.

\begin{table}[h]
\centering\small 
\caption{The misspecification study, 20 datasets per perturbation (100 for the clean test sets). CRPS vs MixSIAPy: difference in mean continuous ranked probability score between MixSIARPFN and MixSIAPy on the datasets where MixSIAPy converged, with paired Bayesian bootstrap 95\% credible intervals (positive: MixSIARPFN less accurate). Change from clean: the difference between that comparison and the same comparison on the clean test set, in percentage points. Coverage: share of 95\% intervals containing the true proportion, all datasets. MixSIARPFN degraded more than MCMC only under heavy-tailed and skewed errors on geese-like data; biased discrimination reduced every method's coverage.}
\label{tab:misspec}
\resizebox{\textwidth}{!}{\input{tables/si_misspec.tex}}
\end{table}

\section*{Appendix S9. What improved amortised accuracy}
\addcontentsline{toc}{section}{Appendix S9. What improved amortised accuracy}

\begin{figure}[h]
\centering 
\includegraphics[width=0.92\textwidth]{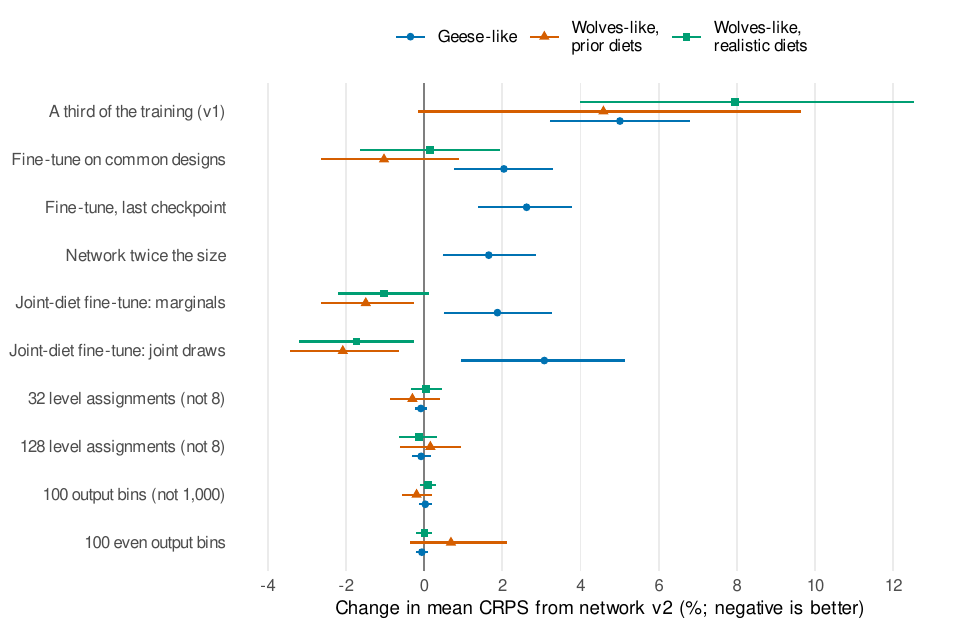}
\caption{What changed amortised accuracy during development. Each point is the paired change in mean continuous ranked probability score (CRPS) from network v2 (each error structure's network after 450,000 training steps, before the joint-diet fine-tune) on the same development datasets, with a Bayesian bootstrap 95\% credible interval; negative is better. Training for a third as long was the largest effect. The joint-diet fine-tune improved the wolves-like designs (Residual~$\times$~Process network) but worsened the geese-like one (Residual only network, not released), and it is confounded with 150,000 extra training steps. These datasets guided the choice of the released networks, so the results are optimistic compared with the held-out test sets.}
\label{fig:ablations}
\end{figure}

On the development datasets, training length mattered most (Fig.~\ref{fig:ablations}). Networks trained for a third as long (150,000 steps instead of 450,000) were 5\% to 8\% less accurate. Validation loss was still falling slowly when training stopped (Fig.~\ref{fig:training}). Further training weighted towards common designs did not improve accuracy. Nor did a network twice the size (34 million parameters). That network was, however, grown from the trained network and fine-tuned for only 150,000 steps. It is therefore not a test of size trained from scratch.

Inference-time averaging over more level assignments made no difference. Nor did a histogram of 100 rather than 1,000 bins. However, 100 evenly spaced bins reduced coverage on the wolves-like data with prior diets to 72\%.

The joint-diet fine-tune of the released Residual~$\times$~Process network made the wolves-like marginals 1\% to 2\% more accurate (Appendix S6). The gain was largest for proportions strongly correlated with another source in the posterior. The fine-tune also added 150,000 training steps, and we did not train a control with the same steps and no joint task. The gain may therefore come from the extra training rather than the joint task. The same fine-tune of the Residual only network, stopped early at 90,000 steps, made the geese-like marginals worse. Its conditional predictions responded too weakly to the sources already drawn, and it was not released.

\section*{Appendix S10. Beyond the training envelope}
\addcontentsline{toc}{section}{Appendix S10. Beyond the training envelope}

The package refuses designs outside the training envelope. The networks, however, have no fixed number of consumers, tracers or sources, because they attend over each as a set. Only the number of levels per factor (at most 20, the pool of level embeddings) and of factors and covariates are fixed by the architecture. We tested whether some limits could become warnings instead of refusals.

We simulated 100 datasets per condition from MixSIAR's model with one fixed effect of four levels, Residual~$\times$~Process error and a uniform prior. The conditions were beyond the envelope (2,000 and 5,000 consumers; 5 and 6 tracers; 9 and 10 sources) and at its edge. We fitted them with the released network, with the envelope check bypassed. We also fitted 10 per condition with MixSIAPy (\texttt{src/python/extrapolation.py}, \texttt{22\_extrapolation.R}).

Calibration held on every axis: the coverage of 95\% intervals was 0.95 to 0.96 throughout. With 5 or 6 tracers the network's CRPS was 6\% to 8\% above MixSIAPy's, as with 4. With 9 or 10 sources it matched MixSIAPy's, although with two tracers those posteriors are close to the prior. With more consumers the network became more conservative than the exact posterior. Its CRPS was 6\%, 9\% and 17\% above MixSIAPy's at 1,000, 2,000 and 5,000 consumers. Its posterior standard deviations were 10\% to 30\% larger, and its posteriors narrowed little beyond 2,000 consumers.

A fit with 5,000 consumers took about 90 seconds and 10~GB of memory. We tested only this design and network. Random effects, covariates, concentration dependence and the other error structures were not tested beyond the envelope.

\begin{figure}[h]
\centering 
\includegraphics[width=0.9\textwidth]{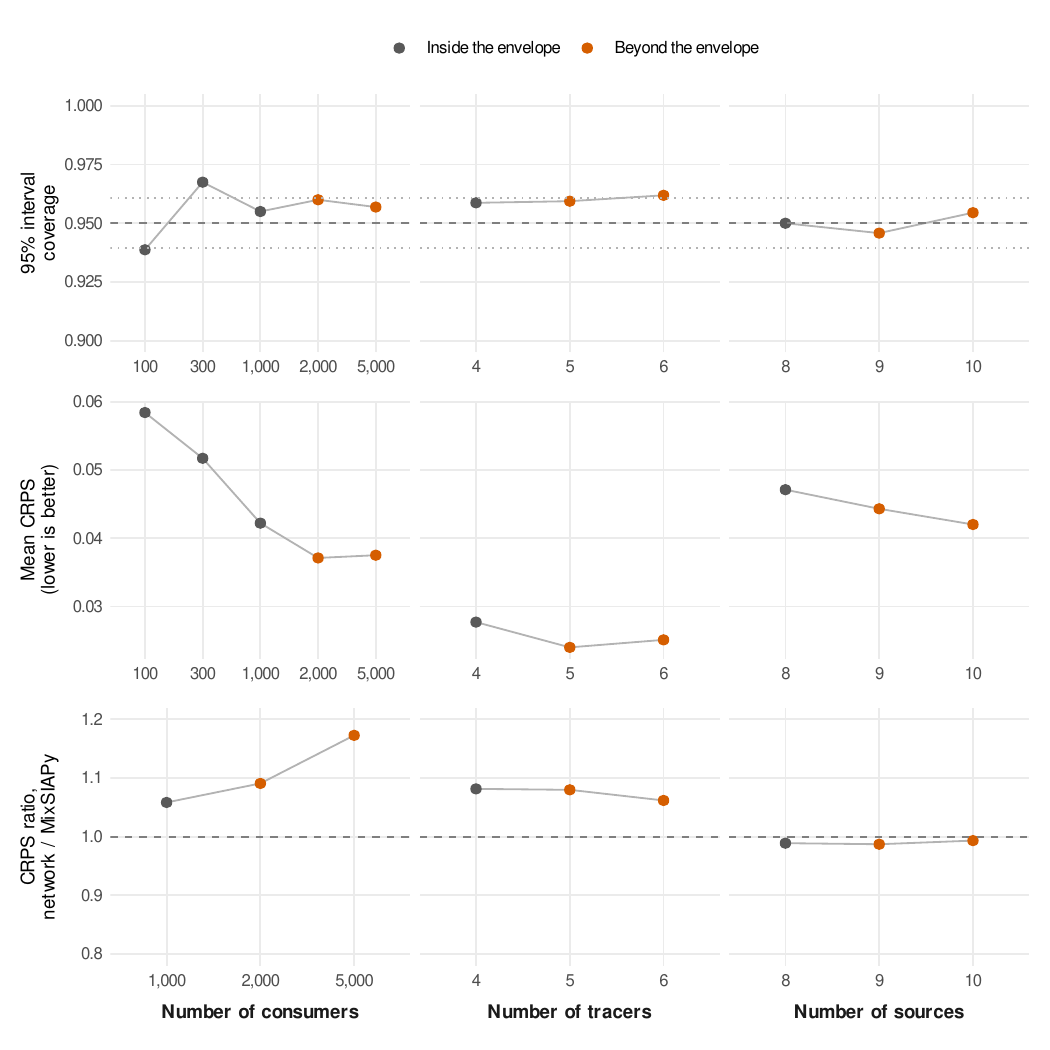}
\caption{Beyond the training envelope: calibration and accuracy of the Residual~$\times$~Process network on simulated datasets with more consumers, tracers or sources than it was trained on (grey: inside the envelope; orange: beyond it). Dashed lines mark nominal coverage and equal accuracy. The networks stayed calibrated on every axis; accuracy relative to MCMC held for more tracers and sources but declined with more consumers, whose extra information the network used only partly.}
\label{fig:extrapolation}
\end{figure}

\end{document}

%% file: tables/table1_heldout.tex
\begin{tabular}{@{}lrlrlll@{}}
\toprule
 & & \multicolumn{2}{c}{Against MixSIAPy, all datasets} & Against converged & & \\
\cmidrule(lr){3-4}
Test design & Proportions & CRPS, \% (95\% CI) & Better, \% & MixSIAR long (datasets) & Coverage & Width, \% \\
\midrule
Geese-like & 3,200 & +3.9 (-0.3, +8.1) & 32 & +3.7 (-0.6, +8.0) (93) & 0.962 (0.954) & +9 \\
Wolves-like, prior diets & 3,600 & +3.9 (+0.9, +7.2) & 38 & +4.3 (+0.7, +8.5) (47) & 0.970 (0.971) & +4 \\
Wolves-like, realistic diets & 3,600 & +2.9 (-0.1, +5.9) & 36 & +3.5 (-0.4, +7.3) (66) & 0.973 (0.968) & +6 \\
Geese-like, Process only & 3,200 & -3.6 (-11.9, +4.8) & 34 & +12.2 (+5.8, +19.4) (54) & 0.943 (0.923) & +20 \\
\bottomrule
\end{tabular}

%% file: tables/si_envelope.tex
\begin{tabular}{@{}ll@{}}
\toprule
Feature & Supported range
\\ \midrule
Sources & 2 to 8 \\
Tracers & 1 to 4 \\
Levels of one effect & 2 to 20 \\
Levels of each of two crossed effects & 2 to 10 \\
Outer levels of a nested pair & 2 to 8 \\
Inner levels per outer level & 1 to 6 (at most 20 in all) \\
Consumers per level & 1 to 300 \\
Consumers in total & 1 to 1,000 (2 or more unless Process only) \\
Source sample size & 2 to 300 \\
Dirichlet prior weight per source & 0.01 to 50 \\
Ratio of source concentrations & up to 100 \\
Continuous covariates & 0 or 1 \\
Effects & none; one fixed or random; two fixed or random, crossed or nested \\
Error structures & Residual * Process; Residual only; Process only \\
\bottomrule
\end{tabular}

%% file: tables/si_training.tex
\begin{tabular}{@{}llrrrr@{}}
\toprule
Network & Stage & Steps & Best step & Best validation loss & GPU-hours
\\ \midrule
Residual * Process & first training & 150,000 & 150,000 & -2.2090 & 33 \\
 & continued & 300,000 & 300,000 & -2.2352 & 66 \\
 & joint-diet fine-tune (released) & 150,000 & 148,000 & -2.2373 & 33 \\
Residual only & first training & 150,000 & 150,000 & -2.7193 & 30 \\
 & continued (released) & 300,000 & 300,000 & -2.7501 & 59 \\
Process only & first training & 150,000 & 150,000 & -3.0990 & 33 \\
 & continued (released) & 300,000 & 298,000 & -3.1289 & 65 \\
\bottomrule
\end{tabular}

%% file: tables/si_calibration.tex
\begin{tabular}{@{}lrrrrrr@{}}
\toprule
Network & 50\% & 90\% & 95\% & PIT KS distance & 95\%, lowest subgroup & 95\%, highest subgroup
\\ \midrule
Residual * Process & 0.500 & 0.900 & 0.950 & 0.0004 & 0.949 & 0.951 \\
Residual only & 0.501 & 0.900 & 0.950 & 0.0012 & 0.949 & 0.951 \\
Process only & 0.500 & 0.900 & 0.950 & 0.0009 & 0.949 & 0.952 \\
\bottomrule
\end{tabular}

%% file: tables/si_sbc.tex
\begin{tabular}{@{}llrrrrrrl@{}}
\toprule
Network & Quantity & Datasets & KS distance & 50\% & 90\% & 95\% & Mean PIT & Inside band
\\ \midrule
Residual * Process & Global diet & 2,000 & 0.005 & 0.50 & 0.90 & 0.95 & 0.50 & yes \\
Residual * Process & Level diet & 1,685 & 0.003 & 0.50 & 0.90 & 0.95 & 0.50 & no \\
Residual * Process & Diet at a covariate value &   715 & 0.011 & 0.49 & 0.90 & 0.95 & 0.50 & yes \\
Residual * Process & Diet of a combination & 1,070 & 0.003 & 0.50 & 0.90 & 0.95 & 0.50 & yes \\
Residual * Process & Random-effect SD & 1,110 & 0.024 & 0.50 & 0.90 & 0.94 & 0.50 & yes \\
Residual * Process & Covariate slope (CLR) &   715 & 0.013 & 0.50 & 0.90 & 0.95 & 0.50 & no \\
Residual * Process & Residual multiplier & 2,000 & 0.014 & 0.51 & 0.90 & 0.95 & 0.50 & yes \\
Residual only & Global diet & 2,000 & 0.005 & 0.50 & 0.90 & 0.95 & 0.50 & no \\
Residual only & Level diet & 1,670 & 0.003 & 0.50 & 0.90 & 0.95 & 0.50 & yes \\
Residual only & Diet at a covariate value &   722 & 0.008 & 0.50 & 0.90 & 0.95 & 0.50 & yes \\
Residual only & Diet of a combination & 1,101 & 0.003 & 0.50 & 0.90 & 0.95 & 0.50 & yes \\
Residual only & Random-effect SD & 1,074 & 0.047 & 0.51 & 0.90 & 0.95 & 0.48 & yes \\
Residual only & Covariate slope (CLR) &   722 & 0.021 & 0.50 & 0.90 & 0.95 & 0.50 & yes \\
Process only & Global diet & 2,000 & 0.009 & 0.51 & 0.90 & 0.95 & 0.50 & yes \\
Process only & Level diet & 1,695 & 0.004 & 0.50 & 0.90 & 0.95 & 0.50 & yes \\
Process only & Diet at a covariate value &   690 & 0.008 & 0.50 & 0.90 & 0.95 & 0.50 & yes \\
Process only & Diet of a combination & 1,090 & 0.008 & 0.50 & 0.90 & 0.95 & 0.50 & yes \\
Process only & Random-effect SD & 1,115 & 0.022 & 0.49 & 0.91 & 0.95 & 0.49 & yes \\
Process only & Covariate slope (CLR) &   690 & 0.015 & 0.50 & 0.91 & 0.96 & 0.50 & yes \\
\bottomrule
\end{tabular}

%% file: tables/si_leakage.tex
\begin{tabular}{@{}p{11cm}l@{}}
\toprule
Check (computed from the files before scoring) & Result
\\ \midrule
Protocol committed before the test sets were generated & pass \\
Test seeds differ from development seeds & pass \\
No test dataset repeats a development dataset's true diets & pass \\
No scores of any method on the test sets yet & pass \\
No script other than the designated fitting and scoring scripts names the test sets & pass \\
Networks trained only on simulations generated during training & pass \\
The networks to be scored are the released v3 files (by checksum) & pass \\
\bottomrule
\end{tabular}

%% file: tables/si_convergence.tex
\begin{tabular}{@{}lrrrrrr@{}}
\toprule
Test design & MixSIAR normal, pre-registered & strict & MixSIAR long, pre-registered & strict & MixSIAPy, pre-registered & strict
\\ \midrule
Geese-like & 83\% & 0\% & 93\% & 0\% & 100\% & 100\% \\
Wolves-like, prior diets & 34\% & 0\% & 47\% & 1\% & 100\% & 89\% \\
Wolves-like, realistic diets & 42\% & 0\% & 66\% & 0\% & 98\% & 95\% \\
Geese-like, Process only & 49\% & 0\% & 54\% & 0\% & 84\% & 63\% \\
\bottomrule
\end{tabular}

%% file: tables/si_selection.tex
\begin{tabular}{@{}lrrrrrr@{}}
\toprule
Test design & Converged & Not converged & CRPS, converged & CRPS, not converged & Normal vs PFN, converged (\%) & Normal vs PFN, not converged (\%)
\\ \midrule
Geese-like & 93 &  7 & 0.055 & 0.055 & -0.1 & -2.6 \\
Wolves-like, prior diets & 47 & 53 & 0.063 & 0.055 & -3.9 & -3.5 \\
Wolves-like, realistic diets & 66 & 34 & 0.066 & 0.058 & -3.7 & -2.6 \\
Geese-like, Process only & 54 & 46 & 0.040 & 0.055 & -10.5 & +18.1 \\
\bottomrule
\end{tabular}

%% file: tables/table2_fast.tex
\begin{tabular}{@{}llllll@{}}
\toprule
Method & Designs fitted & CRPS against MixSIAR long, \% & Coverage & Time per fit & Cores \\
\midrule
MixSIARPFN & 4 of 4 & +3.5 to +12.1 & 0.95 to 0.97 & 0.20 s to 0.76 s & 1 \\
MixSIAPy & 4 of 4 & 0.0 to +0.6 & 0.94 to 0.97 & 4 min to 10 min & 3 \\
cosimmr & 2 of 4 & +77.2 to +219.6 & 0.77 to 0.79 & 17 min to 20 min & 1 \\
MixSIAR normal & 4 of 4 & -0.3 to +3.5 & 0.94 to 0.97 & 18 min to 25 min & 1 \\
MixSIAR long & 4 of 4 & 0 (reference) & 0.94 to 0.97 & 42 min to 60 min & 1 \\
\bottomrule
\end{tabular}

%% file: tables/si_development.tex
\begin{tabular}{@{}lrrrrr@{}}
\toprule
Development design & Datasets & v1 & v2 & q3 & v3 (released)
\\ \midrule
Geese-like & 144 & +10.5 & +5.3 & +7.4 & +5.3 \\
Wolves-like, prior diets &  58 & +8.4 & +3.6 & +2.5 & +1.9 \\
Wolves-like, realistic diets &  90 & +13.8 & +5.4 & +5.6 & +4.1 \\
\bottomrule
\end{tabular}

%% file: tables/si_benchmark_long.tex
{\footnotesize\setlength{\tabcolsep}{4pt}
\begin{longtable}{@{}llrrrp{3.6cm}rrr@{}}
\caption{The 40 benchmark designs on which MixSIAR converged, one simulated dataset each. Error: Residual (only), Process (only) or Res. x Proc. (Residual $\times$ Process). Effects: F, one fixed effect; R, one random effect; FF, FR and RR, two effects (fixed or random) crossed; RR nested, a random effect nested in another; conc., concentration dependence; inform. prior, an informative Dirichlet prior. W1 is the mean Wasserstein-1 distance between the MixSIARPFN and converged MixSIAR posteriors over the design's diet proportions, in units of the proportion; coverage is the share of true diets inside each method's 95\% interval. Thirty-five of the 40 designs had a mean distance below 0.03; the largest, 0.08, was for two crossed random effects.}\label{tab:benchmark}\\
\toprule
Family & Error & Sources & Tracers & Consumers & Effects and options & W1 & \multicolumn{2}{c}{Coverage} \\ \cmidrule(l){8-9} & & & & & & & PFN & MixSIAR \\ \midrule
\endfirsthead
\toprule
Family & Error & Sources & Tracers & Consumers & Effects and options & W1 & \multicolumn{2}{c}{Coverage} \\ \cmidrule(l){8-9} & & & & & & & PFN & MixSIAR \\ \midrule
\endhead
Geese & Residual & 2 & 1 &  20 & F, conc. & 0.011 & 1.00 & 1.00 \\
Geese & Residual & 3 & 2 &  60 & F, conc. & 0.014 & 0.92 & 0.92 \\
Geese & Residual & 3 & 4 & 300 & F, conc. & 0.019 & 1.00 & 1.00 \\
Geese & Residual & 4 & 2 & 250 & F, conc., inform. prior & 0.007 & 1.00 & 1.00 \\
Geese & Residual & 4 & 3 &  30 & F, conc. & 0.007 & 1.00 & 1.00 \\
Geese & Residual & 5 & 2 & 150 & F, conc. & 0.010 & 0.90 & 0.92 \\
Geese & Residual & 5 & 3 & 400 & F, conc., inform. prior & 0.012 & 0.89 & 0.91 \\
Geese & Residual & 6 & 1 &  40 & F, conc., inform. prior & 0.042 & 1.00 & 1.00 \\
Geese & Residual & 7 & 4 & 800 & F, conc. & 0.008 & 0.99 & 0.99 \\
Geese & Residual & 8 & 2 & 100 & F, conc. & 0.007 & 0.94 & 0.94 \\
General & Process & 3 & 1 &  75 & F + covariate & 0.033 & 1.00 & 1.00 \\
General & Process & 4 & 2 &  30 & none & 0.005 & 1.00 & 1.00 \\
General & Process & 4 & 3 &  18 & F, inform. prior & 0.008 & 0.96 & 0.98 \\
General & Res. x Proc. & 2 & 1 &  30 & RR nested, sources by factor & 0.011 & 0.92 & 1.00 \\
General & Res. x Proc. & 2 & 2 & 120 & RR nested, sources by factor, inform. prior & 0.015 & 0.72 & 0.72 \\
General & Res. x Proc. & 3 & 2 & 150 & RR & 0.082 & 1.00 & 0.99 \\
General & Res. x Proc. & 3 & 2 &  66 & RR nested, sources by factor & 0.023 & 0.97 & 1.00 \\
General & Res. x Proc. & 3 & 3 &  60 & RR nested, sources by factor & 0.022 & 1.00 & 1.00 \\
General & Res. x Proc. & 3 & 2 & 300 & RR nested, sources by factor, inform. prior & 0.025 & 1.00 & 1.00 \\
General & Res. x Proc. & 4 & 2 & 120 & FR & 0.032 & 1.00 & 1.00 \\
General & Res. x Proc. & 4 & 3 &  90 & R, conc., sources by factor & 0.012 & 1.00 & 1.00 \\
General & Res. x Proc. & 4 & 2 & 100 & RR nested, sources by factor, inform. prior & 0.024 & 0.96 & 1.00 \\
General & Res. x Proc. & 4 & 4 & 150 & RR nested, sources by factor & 0.013 & 0.98 & 0.98 \\
General & Res. x Proc. & 5 & 2 & 200 & RR nested, sources by factor & 0.032 & 0.99 & 0.99 \\
General & Res. x Proc. & 6 & 1 &  40 & RR nested, sources by factor & 0.011 & 1.00 & 1.00 \\
General & Residual & 3 & 2 &  80 & FF, inform. prior & 0.010 & 1.00 & 1.00 \\
General & Residual & 3 & 1 & 120 & R & 0.012 & 1.00 & 1.00 \\
General & Residual & 5 & 2 & 180 & RR nested, conc. & 0.020 & 0.97 & 0.95 \\
Killer whale & Res. x Proc. & 2 & 1 &   5 & none & 0.004 & 1.00 & 1.00 \\
Killer whale & Res. x Proc. & 2 & 2 &  40 & none & 0.002 & 1.00 & 1.00 \\
Killer whale & Res. x Proc. & 3 & 1 &   2 & none, inform. prior & 0.006 & 1.00 & 1.00 \\
Killer whale & Res. x Proc. & 3 & 2 &  15 & none & 0.007 & 1.00 & 1.00 \\
Killer whale & Res. x Proc. & 4 & 3 &   8 & none & 0.010 & 1.00 & 1.00 \\
Killer whale & Res. x Proc. & 4 & 4 & 120 & none, inform. prior & 0.024 & 1.00 & 1.00 \\
Killer whale & Res. x Proc. & 5 & 2 &  12 & none & 0.002 & 1.00 & 1.00 \\
Killer whale & Res. x Proc. & 5 & 3 &  60 & none & 0.009 & 1.00 & 1.00 \\
Killer whale & Res. x Proc. & 6 & 2 &  25 & none, inform. prior & 0.005 & 0.83 & 0.83 \\
Killer whale & Res. x Proc. & 6 & 4 &   4 & none & 0.005 & 1.00 & 1.00 \\
Killer whale & Res. x Proc. & 8 & 1 &  30 & none & 0.004 & 1.00 & 1.00 \\
Killer whale & Res. x Proc. & 8 & 3 & 300 & none, inform. prior & 0.006 & 1.00 & 0.88 \\
\bottomrule
\end{longtable}}

%% file: tables/si_examples.tex
\begin{tabular}{@{}lrrrlrrrrr@{}}
\toprule
Model & Consumers & Sources & Tracers & Error & Mean abs. diff. & W1 & SD ratio & PFN (s) & MixSIAR (min)
\\ \midrule
Killer whale &  12 & 5 & 2 & Res. x Proc. & 0.023 & 0.024 & 1.09 & 0.04 & 23 \\
Geese & 251 & 4 & 2 & Residual only & 0.077 & 0.080 & 1.38 & 0.77 & 176 \\
Wolves &  66 & 3 & 2 & Res. x Proc. & 0.049 & 0.055 & 1.29 & 0.26 & 377 \\
Alligator, model 1 & 181 & 2 & 2 & Res. x Proc. & 0.003 & 0.008 & 1.25 & 0.09 & 68 \\
Alligator, model 2 & 181 & 2 & 2 & Res. x Proc. & 0.011 & 0.013 & 1.23 & 0.51 & 76 \\
Alligator, model 3 & 181 & 2 & 2 & Res. x Proc. & 0.006 & 0.011 & 1.24 & 0.55 & 73 \\
Alligator, model 4 & 181 & 2 & 2 & Res. x Proc. & 0.009 & 0.014 & 1.28 & 0.57 & 71 \\
Alligator, model 5 & 181 & 2 & 2 & Res. x Proc. & 0.018 & 0.019 & 1.29 & 0.25 & 86 \\
Alligator, model 6 & 181 & 2 & 2 & Res. x Proc. & 0.045 & 0.046 & 1.35 & 0.53 & 84 \\
Alligator, model 7 & 181 & 2 & 2 & Res. x Proc. & 0.021 & 0.023 & 1.18 & 1.75 & 91 \\
Alligator, model 8 & 181 & 2 & 2 & Res. x Proc. & 0.014 & 0.022 & 1.32 & 0.53 & 77 \\
Lake &  21 & 3 & 2 & Residual only & 0.045 & 0.063 & 0.95 & 0.17 & 6 \\
Mantis shrimp &  31 & 6 & 2 & Res. x Proc. & 0.008 & 0.010 & 1.12 & 0.13 & 35 \\
Palmyra &  92 & 3 & 2 & Res. x Proc. & 0.031 & 0.034 & 1.17 & 0.20 & 58 \\
Snail &  58 & 2 & 1 & Res. x Proc. & 0.002 & 0.002 & 1.09 & 0.03 & 2 \\
Storm petrel & 135 & 3 & 2 & Res. x Proc. & 0.010 & 0.011 & 1.12 & 0.28 & 71 \\
\bottomrule
\end{tabular}

%% file: tables/si_ppc.tex
\begin{tabular}{@{}lrrrrr@{}}
\toprule
Data & Datasets & Statistics & Outside 95\% & Outside 99\% & Log density p
\\ \midrule
Geese example & 1 & 40 & 11 & 7 & 0.63 \\
Wolves example & 1 & 40 & 5 & 4 & 0.56 \\
Killer whale example & 1 &  5 & 0 & 0 & 0.65 \\
Geese-like simulations & 6 & 40 & 0 to 2 & 0 & 0.58 to 0.63 \\
Wolves-like simulations & 6 & 40 & 0 to 2 & 0 to 1 & 0.56 to 0.58 \\
\bottomrule
\end{tabular}

%% file: tables/si_joint.tex
\begin{tabular}{@{}lrrrrrr@{}}
\toprule
Development design & Diets & Mean abs. diff. in correlations & Strongest pair r, MixSIAR & ... joint draws & Coverage of sum, joint & ... MixSIAR
\\ \midrule
Geese-like & 1296 & 0.305 & -0.84 & -0.67 & 0.96 & 0.93 \\
Wolves-like, prior diets &  696 & 0.073 & -0.83 & -0.80 & 0.97 & 0.96 \\
Wolves-like, realistic diets & 1080 & 0.082 & -0.83 & -0.80 & 0.98 & 0.96 \\
\bottomrule
\end{tabular}

%% file: tables/si_compute.tex
\begin{tabular}{@{}lrrrr@{}}
\toprule
Item & GPU-hours & Hours & Energy (kWh) & CO2 (kg)
\\ \midrule
Attempts not released (same architecture) & 198 & 110 & 69 & 17.6 \\
Earlier prototypes (single-vignette networks) & 48 & 48 & 17 & 4.3 \\
Released networks (v3) & 320 & 160 & 112 & 28.5 \\
MixSIAR runs (development and benchmark) & 0 & 2448 & 26 & 6.5 \\
MixSIAR runs (examples) & 0 & 18 & 0.19 & 0.049 \\
MixSIAPy and cosimmr (held-out test sets) & 0 & 260 & 3 & 0.7 \\
MixSIAR runs (held-out test sets) & 0 & 693 & 7 & 1.8 \\
MixSIARPFN (held-out test sets) & 0 & 0 & <0.01 & <0.001 \\
Timing benchmark (all methods) & 0 & 34 & 0.35 & 0.089 \\
Posterior predictive checks (MixSIAR) & 0 & 42 & 0.43 & 0.1 \\
Total & 565 &  & 234 & 59.8 \\
\bottomrule
\end{tabular}

%% file: tables/si_misspec.tex
\begin{tabular}{@{}lllllr@{}}
\toprule
Design & Perturbation & MixSIAPy converged & CRPS vs MixSIAPy, \% & Change from clean, points & Coverage, \% (PFN / MixSIAPy / MixSIAR)
\\ \midrule
Geese-like & None (test set) & 100 of 100 & +3.8 (-0.5, +8.1) &  & 96.2 / 95.5 / 94.5 \\
Geese-like & Heavy-tailed errors & 20 of 20 & +13.4 (+3.4, +23.0) & +9.7 (-2.0, +20.5) & 95.0 / 93.6 / 94.9 \\
Geese-like & Skewed errors & 20 of 20 & +11.2 (+4.8, +20.4) & +7.5 (-0.2, +17.2) & 95.3 / 93.5 / 93.9 \\
Geese-like & Outlying consumers & 20 of 20 & +7.5 (-3.2, +18.4) & +3.7 (-8.1, +15.3) & 94.4 / 93.9 / 93.1 \\
Geese-like & Biased discrimination & 20 of 20 & +7.5 (+0.6, +14.2) & +3.7 (-4.5, +12.3) & 88.5 / 86.7 / 85.7 \\
Geese-like & Overstated source precision & 20 of 20 & +4.9 (-5.9, +17.5) & +1.1 (-10.4, +14.4) & 95.1 / 92.5 / 92.4 \\
Wolves-like, realistic & None (test set) & 97 of 100 & +3.5 (+0.7, +6.6) &  & 97.3 / 96.8 / 96.5 \\
Wolves-like, realistic & Heavy-tailed errors & 15 of 20 & +4.5 (+0.2, +9.5) & +0.9 (-4.3, +6.3) & 96.1 / 95.6 / 95.1 \\
Wolves-like, realistic & Skewed errors & 20 of 20 & +3.7 (-0.5, +8.2) & +0.2 (-4.6, +5.0) & 97.2 / 96.7 / 96.8 \\
Wolves-like, realistic & Outlying consumers & 19 of 20 & -2.1 (-9.2, +5.4) & -5.6 (-13.8, +2.8) & 96.8 / 94.3 / 94.9 \\
Wolves-like, realistic & Biased discrimination & 20 of 20 & +3.1 (-3.7, +13.5) & -0.5 (-7.6, +10.1) & 92.9 / 92.6 / 92.5 \\
Wolves-like, realistic & Overstated source precision & 20 of 20 & +3.9 (-2.9, +10.9) & +0.4 (-6.5, +7.8) & 96.5 / 94.0 / 92.8 \\
\bottomrule
\end{tabular}